\documentclass[journal]{IEEEtran}
\usepackage{cite}
\usepackage{graphicx}
\usepackage{hyperref}
\usepackage{amsmath,amssymb,amsfonts}
\usepackage{algorithm}
\usepackage{algpseudocode}
\usepackage{subfig}
\usepackage{hyperref}
\hypersetup{
    colorlinks = true,
    citecolor  = blue,
}

\usepackage{listings}

\lstdefinestyle{mystyle}{
  backgroundcolor=\color{backcolour}, commentstyle=\color{codegreen},
  keywordstyle=\color{magenta},
  numberstyle=\tiny\color{codegray},
  stringstyle=\color{codepurple},
  basicstyle=\fontfamily{pcr}\footnotesize,
  breakatwhitespace=false,         
  breaklines=true,                 
  captionpos=b,                    
  keepspaces=true,                 
  numbers=left,                    
  numbersep=5pt,                  
  showspaces=false,
  showstringspaces=false,
  showtabs=false,                  
  tabsize=2
}

\ifCLASSINFOpdf
\else
\fi

\ifCLASSOPTIONcaptionsoff
 \usepackage[nomarkers]{endfloat}
\let\MYoriglatexcaption\caption
\renewcommand{\caption}[2][\relax]{\MYoriglatexcaption[#2]{#2}}
\fi
\begin{document}
%
\title{Distilling Large Language Models for Network Active Queue Management}
%
%
%

\author{~
        Shiva Raj Pokhrel, Deol Satish, Jonathan Kua~
        and Anwar Walid,~\IEEEmembership{Fellow,~IEEE}
\thanks{S.~R.~Pokhrel, D.~Satish and J.~Kua are with the School of IT, Deakin University, Geelong, Australia. A.~Walid is with Defense Unicorns and Columbia University. corresponding email: shiva.pokhrel@deakin.edu.au }
}

%
%

\markboth{IEEE/ACM Transaction on Networking,~Vol.~14, No.~8, Feburary~2025}%
{Shell \MakeLowercase{\textit{et al.}}: Bare Demo of IEEEtran.cls for IEEE Journals}
%



\maketitle

\begin{abstract}

\color{black}
We propose AQM-LLM, a framework that distills Large Language Models for Active Queue Management in modern networks. Current Deep Learning-based queuing approaches struggle with dynamic network scenarios and require extensive engineering effort. Our approach leverages LLMs' contextual understanding and pattern recognition capabilities to enhance Low Latency, Low Loss, and Scalable Throughput (L4S) architecture with minimal manual intervention. The L4S-LLM implementation incorporates three key innovations: (1) an improved state encoder that processes diverse network metrics into token embeddings, (2) an extended LLM head that generates single-inference congestion decisions, and (3) a data-driven Low-Rank Adaptation scheme that reduces trainable parameters considerably while maintaining very high accuracy. Our open-source FreeBSD-14 implementation demonstrates significant improvements in queue management metrics, with reduced delay variability and enhanced bandwidth utilization across both DCTCP and UDP Prague protocols. Comprehensive evaluations show L4S-LLM effectively balances computational efficiency with network performance, establishing a foundation and novel direction for intelligent, adaptive queuing in next-generation networks.

\end{abstract}


%
\IEEEpeerreviewmaketitle
\newcommand{\R}[1]{\textcolor{blue}{#1}}

\section{Introduction}

\color{black}
Since the 1990s, Active Queue Management (AQM) algorithms have employed mathematical models for optimal rule-based control systems. Early algorithms like Random Early Detection (RED)\cite{floyd1993random} utilized queue length as the primary buffer management metric. Modern AQMs including Controlled Delay (CoDel)\cite{rfc8289}, Proportional Integral Controller Enhanced (PIE)\cite{rfc8033}, and Common Applications Kept Enhanced (CAKE) have transitioned to queue delay-based metrics. This shift enables direct latency control and enhanced network performance. The \textbf{Low Latency, Low Loss, and Scalable Throughput} (L4S) framework [RFC 9330]\cite{rfc9330} exemplifies these advances\footnote{\url{https://www.rfc-editor.org/rfc/rfc9330.html}}. Recent \href{https://www.bell-labs.com/institute/white-papers/l4s-low-latency-low-loss-and-scalable-throughput/}{Bell Labs findings} validate these approaches~\cite{freebsdl4s}.

Rule-based AQMs encounter critical limitations in contemporary heterogeneous networks. WiFi, 5G/6G, optical fiber, and satellite environments exhibit rapid link quality fluctuations. Traffic patterns and interference levels change faster than formulaic adaptation capabilities. Dynamic heterogeneous challenges necessitate complex, computationally intensive rules. Continuous parameter tuning and periodic re-engineering, perhaps by employing machine learning (ML), become mandatory\footnote{\url{https://www.bell-labs.com/white-papers/l4s/}, accessed 15 Aug 2025}\cite{pokhrel2021learning}.

Deep learning provides a paradigm shift toward learning-based AQMs. These systems minimize dependence on predefined mathematical models~\cite{9953949}. Supervised learning handles traffic classification and bandwidth estimation tasks. Reinforcement learning addresses congestion control, adaptive bitrate streaming, and cloud scheduling~\cite{gomez2019intelligent}. Deep neural networks (DNNs) and RL optimize packet traffic prediction in AQM contexts~\cite{freebsdl4s}. Learning-based methods demonstrate superior dynamic adaptation compared to rule-based systems. Data-driven insights enable effective handling of complex scenarios~\cite{freebsdl4s}. Nevertheless, critical challenges impede ML-driven AQM deployment:\\
(1) \textit{Design Complexity:} ML-based AQM model development \textit{demands substantial effort}. Architecture design, hyperparameter optimization, and algorithm selection require extensive experimentation. Computational resource requirements hinder dynamic network deployment.\\
(2) \textit{Generalization Limitations:} ML-based AQMs \textit{struggle with cross-condition generalization}. Models trained on static traffic exhibit poor real-world performance. Reliability remains inferior to traditional rule-based systems.\\
(3) \textit{Retraining Overhead:} Continuous \textit{ML-based AQM retraining} is essential for network evolution adaptation. Diverse, high-quality datasets and extensive preprocessing increase operational costs. Commercial deployment complexity escalates significantly.

\color{black}

\subsection{LLMs over AQM: Challenges \& Distillation}
Leveraging Large Language Models (LLMs) like GPT, Opt, and Llama~\cite{touvron2023llama, guo2025deepseek} into AQM can address the high costs of developing task-specific DNNs for solving networking challenges~\cite{wu2024netllm, pokhrel2024large}. Pretrained on vast text datasets, LLMs excel in contextual understanding, reasoning, and generalization~\cite{guo2025deepseek} making them ideal for tasks like AQM. Their attention mechanism enables dynamic analysis of traffic parameters and dependencies, allowing proactive congestion management and adaptability to evolving network conditions. The distillation of LLM for AQM offers a universal framework to improve the Internet with minimal modifications, which can perhaps be achieved by adopting new RL ideas for the distillation of knowledge~\cite{guo2025deepseek, 10271124}. However, key challenges include:\\
(1) \textit{Diverse Inputs}: AQM data includes time series (e.g., queue delay) and scalar values (e.g., burst allowance), requiring specialized encoders.\\
(2) \textit{Inference Latency}: Token-by-token output generation causes delays, disrupting AQM’s rapid update cycles (45 ms). Hallucinated outputs further exacerbate the delays.\\
(3) \textit{Distillation Costs}: Distilling LLMs for AQM requires time-consuming fine-tuning, especially in decision-making tasks using reinforcement learning, and demands significant computational resources for large models. While adopting LLMs can unify and enhance AQM, the above mentioned advances must be developed for effective AQM-LLM design and deployment.

\color{black}
\subsection{Our Design Details of L4S-LLM and Contributions}

We propose AQM-LLM and develop and implement L4S-LLM as a proof of concept. L4S-LLM is an adapted LLM framework developed for efficient congestion prevention on the top of L4S. L4S-LLM leverages a foundational LLM model for decision-making in the L4S queues with minimal manual modifications, by incorporating three key modules:\\
    (1) \textit{State Encoder:} A multifaceted encoder processes diverse network data (e.g., queue delay, packet drops) into token-like embeddings, enabling the LLM to understand and utilize network metrics effectively.\\
   (2) \textit{ L4S-LLM Head:} Replacing the default language model head, the L4S-LLM head maps LLM outputs directly to congestion prevention actions (enqueue, drop, or mark) in a single inference, reducing response time and preventing hallucinations.\\
    (3) \textit{Data-Driven Low-Rank L4S Adaptation (proposed LoRA):} This module finetunes the LLM using RL with distillation over the pre-collected data from existing algorithms, eliminating lengthy real-time interactions. By employing Low-Rank Adaptation (LoRA), finetuning costs are drastically reduced, with a 64\% decrease in GPU memory usage and a 15.1\% reduction in training time.

L4S-LLM addresses low-latency needs by proactively identifying traffic patterns and taking optimal congestion-prevention actions. Its strong generalization capability ensures high performance on unseen network scenarios.

 Our key contributions in this paper are as follows.
\begin{enumerate}
    \item Identification of key challenges and advantages of using LLMs for network queue management.
    \item Development of L4S-LLM with a multi-faceted encoder, specialized L4S-LLM head, and the proposed LoRA scheme for improved RL-based distillation and finetuning of the L4S-LLM.
    \item Evaluation demonstrating improved queue management, congestion mitigation, and latency reduction across unseen data and network environments.
\end{enumerate}  

\textit{L4S-LLM Head Design \& Development.} Conventional LLM heads rely on iterative multiround inference to generate valid responses, resulting in substantial latency. To overcome this limitation, we propose the novel L4S-LLM head, which generates a valid response in a single inference round by predicting a probability distribution over possible actions. This design streamlines the response generation process, significantly accelerating inference. Drawing inspiration from \textit{speculative decoding}, the L4S-LLM head advances beyond sequential token-by-token generation by simultaneously evaluating multiple potential outcomes, enhancing both computational efficiency and scalability (without degrading any accuracy). Furthermore, L4S-LLM offers a robust framework to improve AQM performance while reducing manual effort and resource consumption. 

\begin{figure}[t]
\centerline{\includegraphics[width=0.7\linewidth]{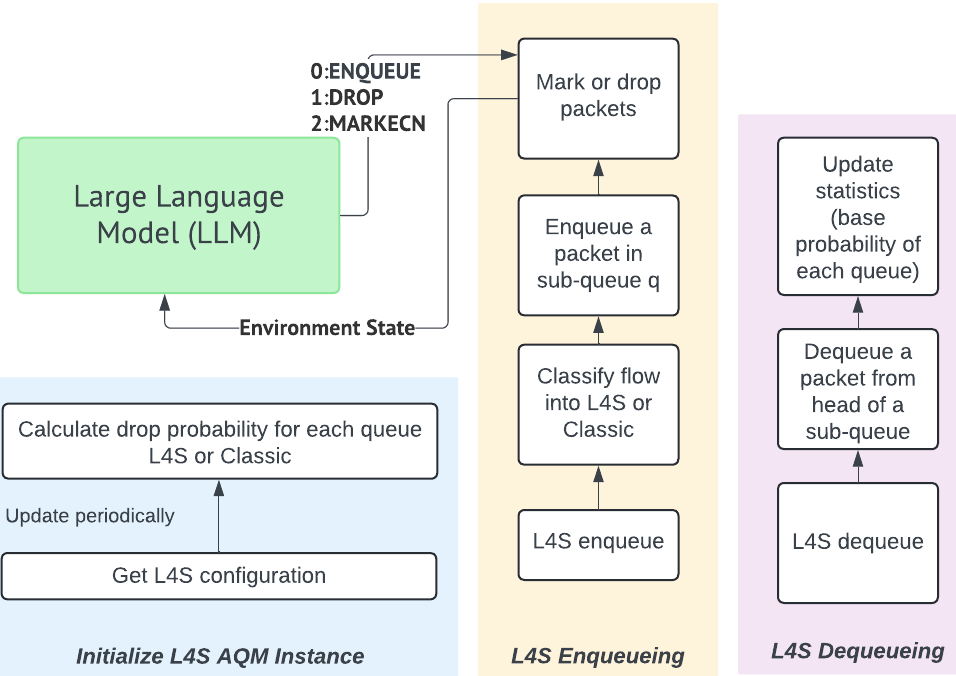}}
 \caption{L4S-LLM Coupling: Interaction between different components of L4S and LLM, illustrating how the model integrates with the existing network stack.}
\label{fig:l4s_llm_coupling}
\end{figure}

Figure~\ref{fig:l4s_llm_coupling} demonstrates the integration of LLM intelligence with L4S network stack. Observe in Figure~\ref{fig:l4s_llm_coupling} how LLM components analyze real-time network telemetry and traffic patterns to dynamically optimize L4S parameters, including dual-queue AQM management, ECN marking thresholds, and congestion control algorithms. A bidirectional feedback loop enables LLM to process network performance data while simultaneously providing intelligent parameter adjustments and predictive congestion management to the existing L4S stack, resulting in enhanced network efficiency and reduced latency through LLM-driven L4S.
\textit{Temporal Modeling Framework.} Our temporal modeling framework employs a dual-stage architecture to capture queue dynamics across multiple timescales. The initial stage utilizes 1D convolutional neural networks (Algorithm~\ref{algo:state_encoder}) with varying kernel sizes to extract multi-scale temporal features from sequential network metrics, providing efficient feature extraction before transformer processing.

The transformer component employs self-attention across an empirically optimized context window of network activity. Ablation studies revealed this configuration maximizes performance while maintaining generalizability. Shorter context windows exhibited performance degradation, while extended windows showed no significant improvement and increased computational complexity. To mitigate temporal overfitting, we implemented three regularization techniques:
\begin{enumerate}
    \item {Protocol-diverse training corpus:} Incorporating heterogeneous network traces from multiple congestion control algorithms
    \item {Stochastic temporal augmentation:} Applying random offsets and noise to temporal sequences
    \item {Temporal dropout:} Implementing random masking of timestep information during training
\end{enumerate}
Our findings highlight the complementary strengths of learning-based and classical approaches. While L4S-LLM offers superior performance in most scenarios, classical algorithms provide valuable safety guarantees and interpretability that inform our ongoing work to combine these strengths through formal verification and explicit incorporation of classical control principles. Empirical validation demonstrates the efficacy of this approach, with the model maintaining high accuracy on previously unseen network conditions. The data-driven reinforcement learning pipeline optimizes for discounted cumulative returns rather than immediate rewards, improving long-horizon performance compared to myopic optimization.

 \textit{Open Source FreeBSD Implementation.}\footnote{https://github.com/MPTCP-FreeBSD/L4S-LLM} We establish an open-source experimental platform to implement and investigate new stacks of \href{https://github.com/MPTCP-FreeBSD/FreeBSD-L4S-Experiments}{L4S within FreeBSD-14}, allowing integration with LLM models. Independent and interoperable implementations are critical for the recognition of emerging LLM-based AQMs by the Internet Engineering Task Force (IETF). In this work, we design and implement \href{https://github.com/MPTCP-FreeBSD/L4S-LLM}{\textit{L4S-AQM framework on FreeBSD-14}}, providing publicly accessible user and kernel modules to facilitate the testing and experimentation of AQM-LLM by the broader network and Internet research community.

\color{black}

\subsection{Literature}~\label{sec:background}
Networks often suffer from `bufferbloat' \cite{9399631,gettys2011bufferbloat}, where excessive packet buffering leads to high latency and jitter. To mitigate this, AQM schemes were developed, which detect congestion and either drop packets or use ECN to signal TCP congestion control algorithms to adjust the congestion window. Rule-based AQMs like CoDel \cite{nichols2012controlling}, PIE \cite{rfc8033}, and CAKE depend on manually created rules, but these static approaches struggle to adapt to dynamic network traffic patterns.

Learning-based AQM algorithms address these limitations by leveraging deep learning techniques to automate decision-making. Early works, such as PFED \cite{gao2005pfed}, forecast traffic using MMSE prediction to regulate queue length and penalize misbehaving flows. Advanced approaches include ECN-based algorithms using LSTM architectures for traffic prediction \cite{gomez2019intelligent}, QP-AQM for adaptive queue management with Q-learning traffic predictors \cite{liu2022active}, and Deep Q-Network-based AQMs for intelligent packet dropping and differential QoS \cite{kim2021deep}.

While these methods improve adaptability, they face challenges like high engineering effort, computational cost, and resource-intensive training. Their reliance on continuous learning limits generalization across diverse network environments, hindering practical deployment in real-world settings. LLMs \cite{vaswani2017attention} like ChatGPT, PaLM, Llama2, and OPT are advanced deep neural networks built on the Transformer architecture, with widespread applications across diverse domains. LLMs process input and output as token sequences, converting text into embeddings and predicting subsequent tokens using an auto-regressive mechanism. The self-attention mechanism allows LLMs to focus on relevant parts of the input, enabling them to handle complex language patterns and nuances effectively. Nevertheless, the LLM Driven networking has potential in several networking aspects including network function virtualization and large scale simulations \cite{8811504, 8565965, huang2025large}.

There are some notable approaches for adapting LLMs to networking domains~\cite{saad2025agi, zou2024genainet}. Supervised Fine-Tuning (SFT)~\cite{chiang2023vicuna} enables efficient domain adaptation with minimal data while preserving instruction-following capabilities, but exhibits safety degradation that increases with dataset size and is particularly sensitive to structured formats like lists and tables \cite{djuhera2025safecomm}. Continual Pre-Training (CPT)~\cite{maatouk2024tele} provides deeper domain knowledge integration and superior performance on complex tasks, but suffers from severe safety degradation when not followed by safety-focused instruction tuning \cite{maatouk2024tele, nikbakht2024tspec}. To address these safety concerns, several methods have been proposed that maintain safety guardrails while preserving task performance. These include SafeInstruct \cite{bianchi2024}, which interleaves safety-aligned QA pairs during fine-tuning; SafeLoRA \cite{hsu2024}, which selectively adapts only layers exhibiting harmful behavior; and SafeMERGE \cite{djuhera2025merge}, which merges fine-tuned layers with those of known safe models. 

\section{Research Design} \color{black}
The L4S architecture, defined in RFC 9330 \cite{rfc9330}, enhances QoE and QoS by integrating ECN with advanced TCP and AQM techniques. Central to this is the Dual-Queue Coupled AQM\footnote{https://www.rfc-editor.org/rfc/rfc9332.html} [RFC 9332], which prioritizes latency-sensitive L4S flows over traditional Classic queues, reducing delay at the cost of minor performance trade-offs for Classic flows. Complementing this, the Prague TCP variant works synergistically with DualPI2 and ECN to minimize packet loss and latency while maintaining high throughput.

DualPI2 uses ECN (RFC 3168 \cite{rfc3168}) to mark packets during congestion, resorting to packet drops only under severe congestion. ECN leverages two bits in the IP header's DiffServ field to signal congestion with four states. Table~\ref{tab:ecntable} shows the details. Non-ECT (00) indicates endpoints don't support ECN, so routers drop packets during congestion. ECT(1) (01) and ECT(0) (10) show endpoints are ECN-capable, allowing routers to mark packets instead of dropping them. CE (11) indicates the packet experienced congestion and is set by routers to notify the receiving endpoint.\footnote{In our \href{ https://man.freebsd.org/cgi/man.cgi?query=tcp}{FreeBSD} implementation, we have configured ECN through the \textit{net.inet.tcp.ecn.enable} sysctl parameter. The default setting activates ECN solely for incoming connections requesting it, but can be modified to apply to all connections or turned off completely.} 

\begin{figure*}[htbp]
    \centering
    \includegraphics[width=0.689\textwidth]{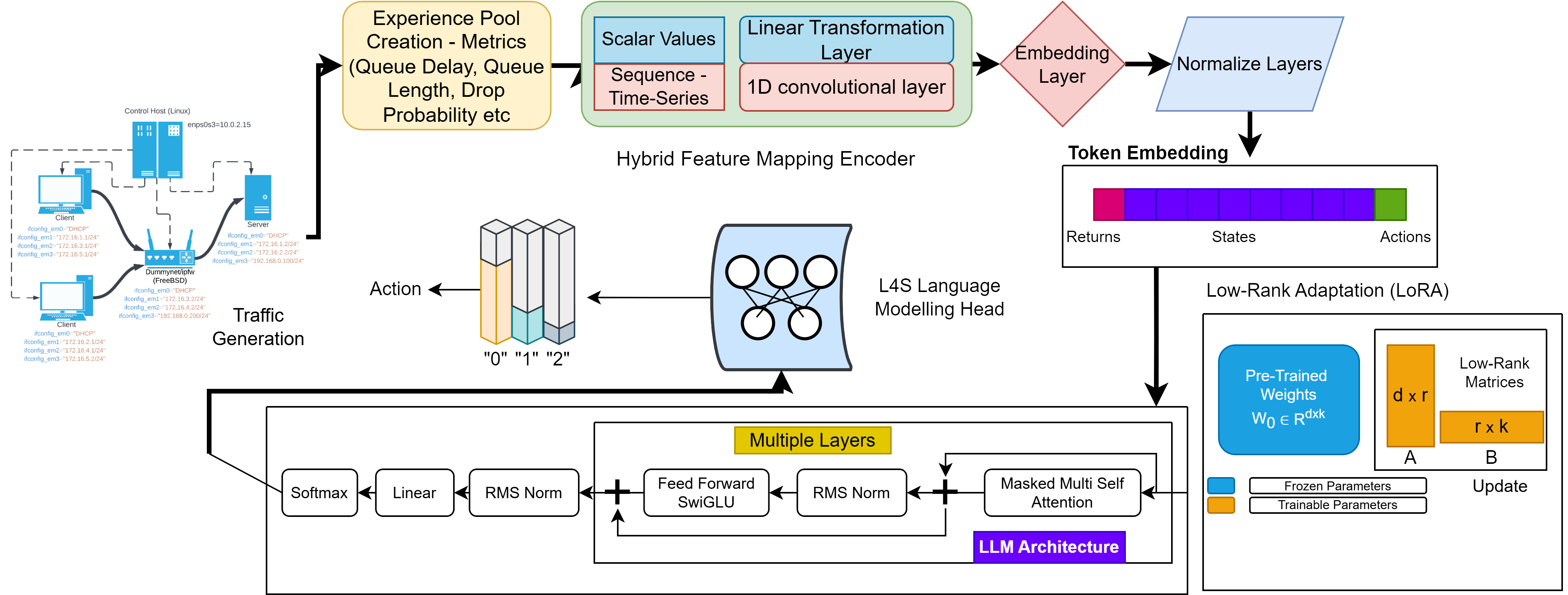}
    \caption{L4S-LLM Architecture: The complete pipeline from raw network metrics to AQM decisions, showing the data flow through the state encoder, transformer model, and L4S-LLM head.}
    \label{fig:l4s_llm_architecture}
\end{figure*}

Upon receiving CE-marked packets, the receiver informs the sender via an ECE flag, prompting the sender to adjust congestion control and respond with a CWR flag to acknowledge the notification. This mechanism mitigates delays, reduces packet loss, and minimizes retransmission, leading to lower latency, improved throughput, and decreased head-of-line blocking \cite{rfc8087}. An abstract view of our L4S-LLM Architecture with complete pipeline from raw network metrics to AQM decisions, are shown in Figure~\ref{fig:l4s_llm_architecture} with the data flow through the state encoder, transformer model, and L4S-LLM head. We discuss the components of the L4S-LLM Architecture in details as follows.

\begin{table}[h]
    \centering
    \caption{ECN Codepoints}
    \begin{tabular}{c|c|c}
        \hline
        \textbf{Codepoint} & \textbf{Binary} & \textbf{Description} \\
        \hline
        00 & 00 00 & Non-ECT (Not ECN-Capable Transport) \\
        \hline
        01 & 00 01 & ECT(0) (ECN-Capable Transport) \\
        \hline
        10 & 00 10 & ECT(1) (ECN-Capable Transport) \\
        \hline
        11 & 00 11 & CE (Congestion Experienced) \\
        \hline
    \end{tabular}
    \label{tab:ecntable}
\end{table}

\subsection{Kernel Log Collection and Format}

\begin{figure}[h!]
    \centering
    \includegraphics[width=1\linewidth]{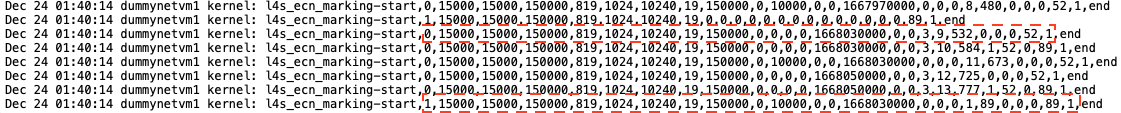}
    \caption{Snapshot of Raw Kernel Log}
    \label{fig:kernellog}
\end{figure}

Our approach begins with collecting raw network metrics directly from the FreeBSD kernel's DualPI2 implementation. These logs capture the complete state of the AQM system at each decision point. Figure~\ref{fig:kernellog} shows an example of the raw kernel log in the format (observe the values in the highlighted dashed rectangles): {\it{queue type, qdelay reference, tupdate, max burst, max ecn threshold, alpha coefficient, beta coefficient, flags, burst allowance, drop probability, 
        current queue delay, previous queue delay, accumulated probability, measurement start time, 
        average dequeue time, dequeue count, status flags, total packets, total bytes, 
        queue length, length in bytes, total drops, packet length, dequeue action}}.

\begin{figure}[h!]
    \centering
    \includegraphics[width=0.5\linewidth]{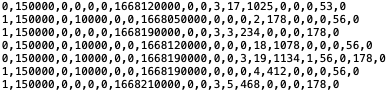}
    \caption{Snapshot of our processed experience pool}
    \label{fig:exp_pool}
\end{figure}
Each log entry represents a single packet processing event with timestamps and key metrics. The final field in the dashed rectangle, \texttt{action}, indicates the decision made by the original DualPI2 algorithm: 0 for enqueue, 1 for drop, and 2 for ECN marking. See Table~\ref{tab:ecntable} for details. These logs are collected using a custom kernel module seamlessly that hooks into the DualPI2 decision path without affecting normal operation.

\subsection{Experience Pool Generation}

The raw kernel logs are processed into structured experience pools for training and evaluating our L4S-LLM model. Each entry contains a state vector of 8 network condition features, an action decision (0: enqueue, 1: drop, 2: ECN mark), a computed reward (balancing throughput and latency), and a Boolean flag refer to as \textit{Done} indicating episode termination (0 except at transmission end). Figure~\ref{fig:exp_pool} shows the snapshot of the processed experience pool. The experience pools are stored in a structured JSON format to facilitate efficient loading during model training.

We demonstrate the L4s-LLM pipeline overview from data collection to LLM head in Table~\ref{tab:l4sllm-stages}. Table~\ref{tab:dualpi2-kernel-metrics} displays all metric data obtained from the kernel along with their descriptions. Custom logging functions were implemented within the kernel to accurately capture data for analysis, offline data cleaning, and preparation for generating the LLM experience pool.
\begin{table*}[!t]
\centering
\caption{Pipeline Stages and Overview}
\label{tab:l4sllm-stages}

\begin{tabular}{l|p{5cm}|p{6.5cm}}
\hline
\textbf{Stage} & \textbf{Description} & \textbf{Example} \\ \hline

Kernel Log Collection & 
Real-time AQM metrics collected from kernel-level code during packet processing & 
Queue Type = 1, Queue Delay = 50ms, Drop Prob = 0.02 \\ \hline

Experience Pool Generation & 
Kernel logs transformed into training/evaluation pools with state, action, reward, and done flags & 
State: $[f_1, f_2, \ldots, f_n]$, Action: $\{0,1,2\}$, Reward: throughput, Done: $\{0,1\}$ \\ \hline

Tokenization & 
Network data converted into discrete tokens for LLM processing & 
Network state $\rightarrow$ [queue\_delay, drop\_prob, pkt\_len, ...] \\ \hline

Embedding & 
Tokens mapped to high-dimensional vectors matching LLM input space & 
queue\_delay $\rightarrow [0.2, 0.7, 0.1, \ldots]$, action $\rightarrow [0.3, 0.1, -0.1, \ldots]$ \\ \hline

RL Adaptation & 
L4S-LLM fine-tuned using rewards to optimize AQM decisions & 
$s_t = (qdelay_t, pdrop_t, lenb_t)$, $R_t = \frac{pkt\_len_t}{qdelay_{t + 1}}$ \\ \hline

LoRA & 
Parameter-efficient training using low-rank matrices while freezing LLM weights & 
$\Delta W = A \cdot B$, where $A, B$ are low-rank matrices \\ \hline

L4S Head & 
Maps LLM outputs to congestion actions (enqueue, drop, mark) & 
$Action = \arg\max(P(enqueue), P(drop), P(mark))$ \\ \hline

\end{tabular}
\end{table*}

\begin{table}[!t]
\centering
\caption{Collected Kernel-Level Metrics from FreeBSD DualPI2 AQM}
\label{tab:dualpi2-kernel-metrics}

\subfloat[Kernel logs utilized by L4S-LLM]{
\begin{tabular}{l|p{5cm}}
\hline
\textbf{Variable} & \textbf{Definition} \\
\hline
queue\_type & Queue type identifier (0: Classic, 1: L4S) \\
\hline
burst\_allowance & Maximum allowed burst size before congestion control activation \\
\hline
drop\_probability & Probability of dropping a packet during congestion \\
\hline
current\_queue\_delay & Current queue delay experienced by packet \\
\hline
accumulated\_probability & Accumulated probability for drop decisions \\
\hline
length\_in\_bytes & Current queue length in bytes \\
\hline
packet\_length & Length of current processed packet \\
\hline
\end{tabular}
}

\vspace{1em}

\subfloat[Non-utilized Kernel log parameters]{
\begin{tabular}{l|p{5cm}}
\hline
\textbf{Variable} & \textbf{Definition} \\
\hline
qdelay\_reference & Target queue delay reference for AQM algorithms \\
\hline
tupdate & Update interval for queue state variables \\
\hline
max\_burst & Maximum allowable burst duration \\
\hline
max\_ecn\_threshold & Upper ECN marking threshold \\
\hline
alpha\_coefficient & PI controller proportional gain \\
\hline
beta\_coefficient & PI controller integral gain \\
\hline
flags & Internal AQM status flags \\
\hline
previous\_queue\_delay & Previous measurement cycle queue delay \\
\hline
measurement\_start\_time & Current measurement interval timestamp \\
\hline
average\_dequeue\_time & Mean packet dequeue time \\
\hline
dequeue\_count & Packets dequeued in current interval \\
\hline
status\_flags & Operational state indicators \\
\hline
total\_packets & Total packets processed \\
\hline
total\_bytes & Total bytes processed \\
\hline
queue\_length & Current queue length in packets \\
\hline
total\_drops & Total packet drops in interval \\
\hline
dequeue\_action & Decision (0:enqueue, 1:drop, 2:ECN mark) \\
\hline
\end{tabular}
}
\end{table}


\subsection{Tokenization and Embedding Process}

Unlike traditional NLP applications where text is tokenized into discrete units, our network metrics require specialized processing to be compatible with LLM architectures. We implement a custom tokenization process that converts numerical network metrics into a format suitable for transformer models.

We present a novel approach to network congestion control by adapting transformer-based language models for sequential decision-making in L4S environments. The architecture fundamentally reframes network state transitions as a sequence modeling problem, leveraging the autoregressive capabilities of pre-trained LLMs. Our tokenization process transforms heterogeneous network telemetry data into a unified sequential representation suitable for transformer processing. Rather than traditional linguistic tokens, the system constructs sequences from temporal network observations, where each timestep $t$ generates a multi-modal token comprising: (1) return-to-go $R_t$ representing cumulative future rewards, (2) decomposed state features $\mathbf{s}_t = [s_{t,1}, s_{t,2}, \ldots, s_{t,7}]$ capturing network conditions, and (3) action $a_t \in \{0,1,2\}$ representing packet handling decisions. 

The sequence construction follows the following pattern:
\begin{equation}
\mathcal{S} = [R_t, s_{t,1}, s_{t,2}, \ldots, a_t, R_{t+1}, s_{t+1,1}, \ldots, a_{t+1}, \ldots]
\end{equation}
which creates a causal dependency structure where states predict subsequent actions, aligning with the autoregressive nature of transformer architectures.

Furthermore, the proposed embedding mechanism employs separate linear projection layers for each modality to map heterogeneous input spaces into a unified $d$-dimensional embedding space matching the pre-trained LLM's hidden size. State features undergo domain-specific transformations through dedicated embedding matrices $\mathbf{W}_{\text{state},i} \in \mathbb{R}^{d \times d_{\text{state}}}$ for $i \in \{1,\ldots,7\}$, while actions and returns are projected via $\mathbf{W}_{\text{action}} \in \mathbb{R}^{d \times 1}$ and $\mathbf{W}_{\text{return}} \in \mathbb{R}^{d \times 1}$ respectively. 

Temporal information is incorporated through learned positional embeddings $\mathbf{W}_{\text{time}} \in \mathbb{R}^{(T+1) \times d}$, which are additively combined with all modality embeddings
\begin{equation}
\text{embed}(\mathbf{x}_t) = \mathbf{W}_{\text{modality}}(\mathbf{x}_t) + \mathbf{W}_{\text{time}}(t)
\end{equation}
where $\mathbf{x}_t$ represents the input at timestep $t$.

The concatenated embedding sequence undergoes layer normalization before being fed into the transformer backbone, bypassing traditional tokenization by directly providing input embeddings rather than discrete tokens. This preserves the continuous nature of network telemetry while leveraging pre-trained attention patterns. The transformer processes sequences through multi-head self-attention mechanisms, capturing complex temporal dependencies and cross-modal interactions between network states, actions, and rewards. The attention mechanism computes weighted representations by calculating similarity scores between query, key, and value matrices, applying softmax normalization, and producing context-aware embeddings. The final hidden states from state embedding positions are extracted and passed through a linear classification head to generate action logits, treating congestion control as a sequence-to-sequence mapping where network history predicts optimal packet handling decisions.

\subsection{Architecture of AQM-LLM}
\begin{algorithm}[t]
\caption{L4S-LLM State Encoder}
\label{algo:state_encoder}
\begin{algorithmic}[1]
\State \textbf{Input:} State tensor \texttt{state} with shape \texttt{(batch\_size, seq\_len, 8, 1)}
\State \textbf{Output:} Encoded feature tensors for each state component
\State Initialize layers:
\State \quad \text{$fc_1$, ..., $fc_n$} as \text{Linear(1,feature\_dim)} for scalar data \label{line:linear_encoder}
\State \quad \text{$conv_1$, ..., $conv_n$} as \text{Conv1d(1, feature\_dim, conv\_size)} for sequence data \label{line:cnn_encoder}
\State \quad \text{$embed\_feature$, ..., $embed\_feature\_n$ as} 
\hspace*{1em} \text{Linear(feature\_dim, plm\_embed\_size) where} 
\hspace*{1em} \text{plm\_embed\_size depends on LLM token size.}
\For{each feature $i$ from 1 to n}
    \If{feature $i$ is scalar data}
        \State Extract \text{feature\_i} from \text{state[:, i-1:i]}
        \State Apply \text{fc\_i(feature\_i)} to encode the feature
    \Else
        \State Extract \text{feature\_i} from \text{state[:, i-1:i]}
        \State Apply \text{conv\_i(feature\_i)} to encode the feature
        \State 
    \EndIf
    \State Apply \text{embed\_i(feature\_i)} to encode the feature \label{line:embedding_layer}
    \Statex \hspace*{1em} into its embedded representation \Comment{Embedding Layer} 
    \State \text{Reshape encoded feature to \text{(batch\_size, seq\_len,-1)}}
\EndFor
\State
\Return \text{Encoded features: {features1, features2, ..., features8}}
\end{algorithmic}
\end{algorithm}
\subsubsection{L4S-LLM State Encoder}

\begin{figure}[h!]
    \centering
    \includegraphics[width=0.5068\linewidth]{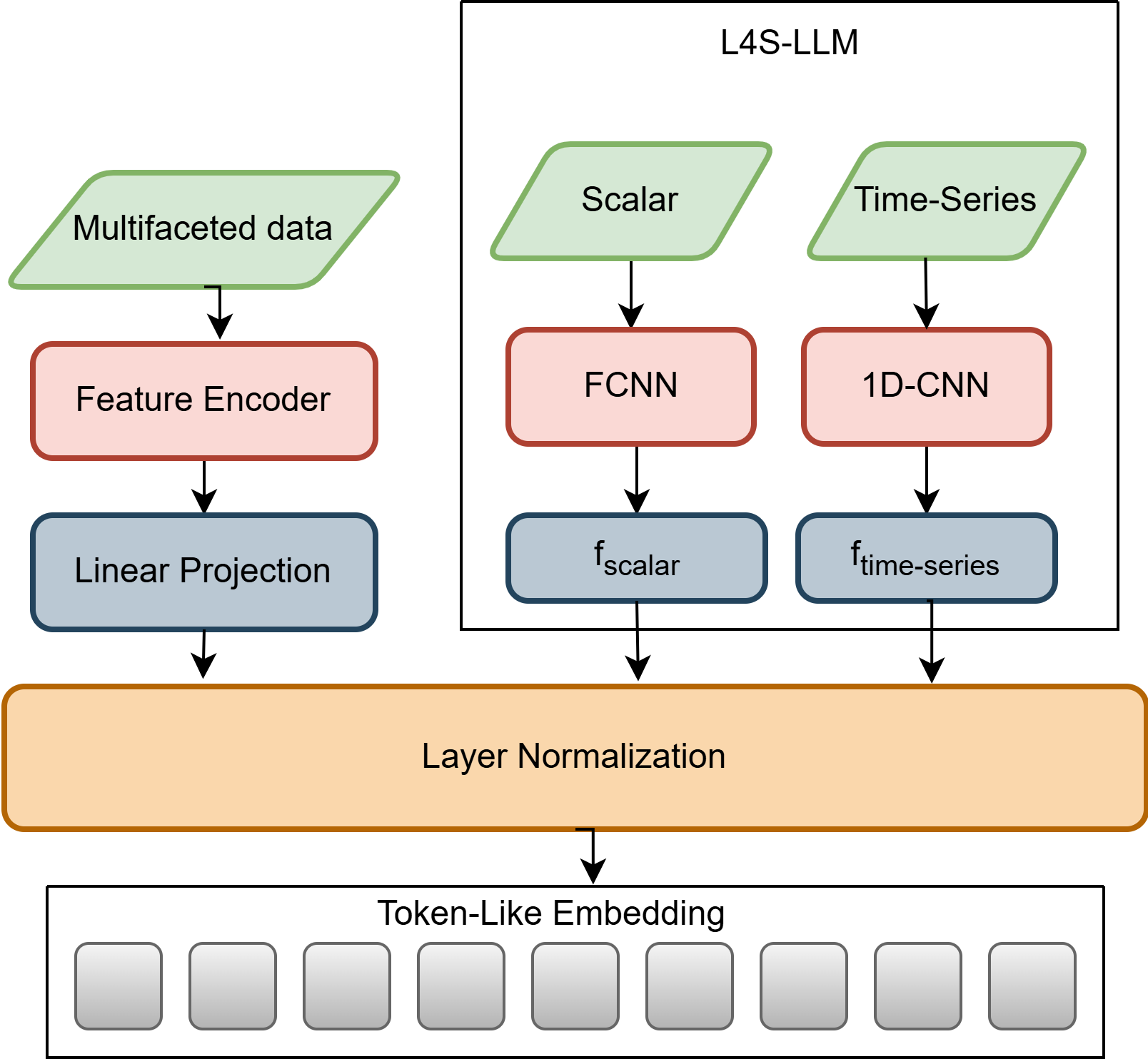}
    \caption{L4S-LLM Encoder: Detailed architecture showing how different types of network metrics are processed through specialized encoding paths.}
    \label{fig:l4s_llm_encoder}
\end{figure}

The L4S-LLM State Encoder transforms raw network metrics into a format compatible with transformer models. As shown in Figure~\ref{fig:l4s_llm_encoder}, it consists of two main components:

\begin{enumerate}
    \item \textbf{Feature Encoder}: Processes different types of network metrics through specialized paths
    \item \textbf{Linear Projector}: Maps extracted features to the token space required by the LLM
\end{enumerate}

Algorithm~\ref{algo:state_encoder} details the encoding process. For scalar metrics like queue type or drop probability, fully connected layers (line~\ref{line:linear_encoder}, Algorithm~\ref{algo:state_encoder}) extract meaningful representations. For time-series data like queue delay trends, 1D CNNs (line~\ref{line:cnn_encoder}, Algorithm~\ref{algo:state_encoder}) capture temporal patterns with kernel sizes $k \in \{3,5,7\}$ to extract multi-scale features.

The extracted features are then projected into the LLM's token space using trainable embedding layers (line~\ref{line:embedding_layer}, Algorithm~\ref{algo:state_encoder}). For example, with Llama2, features are mapped to 4096-dimensional vectors compatible with the model's input requirements.

\subsubsection{L4S Language Modelling Head}

\begin{figure}[h!]
\centerline{\includegraphics[width=0.6\linewidth]{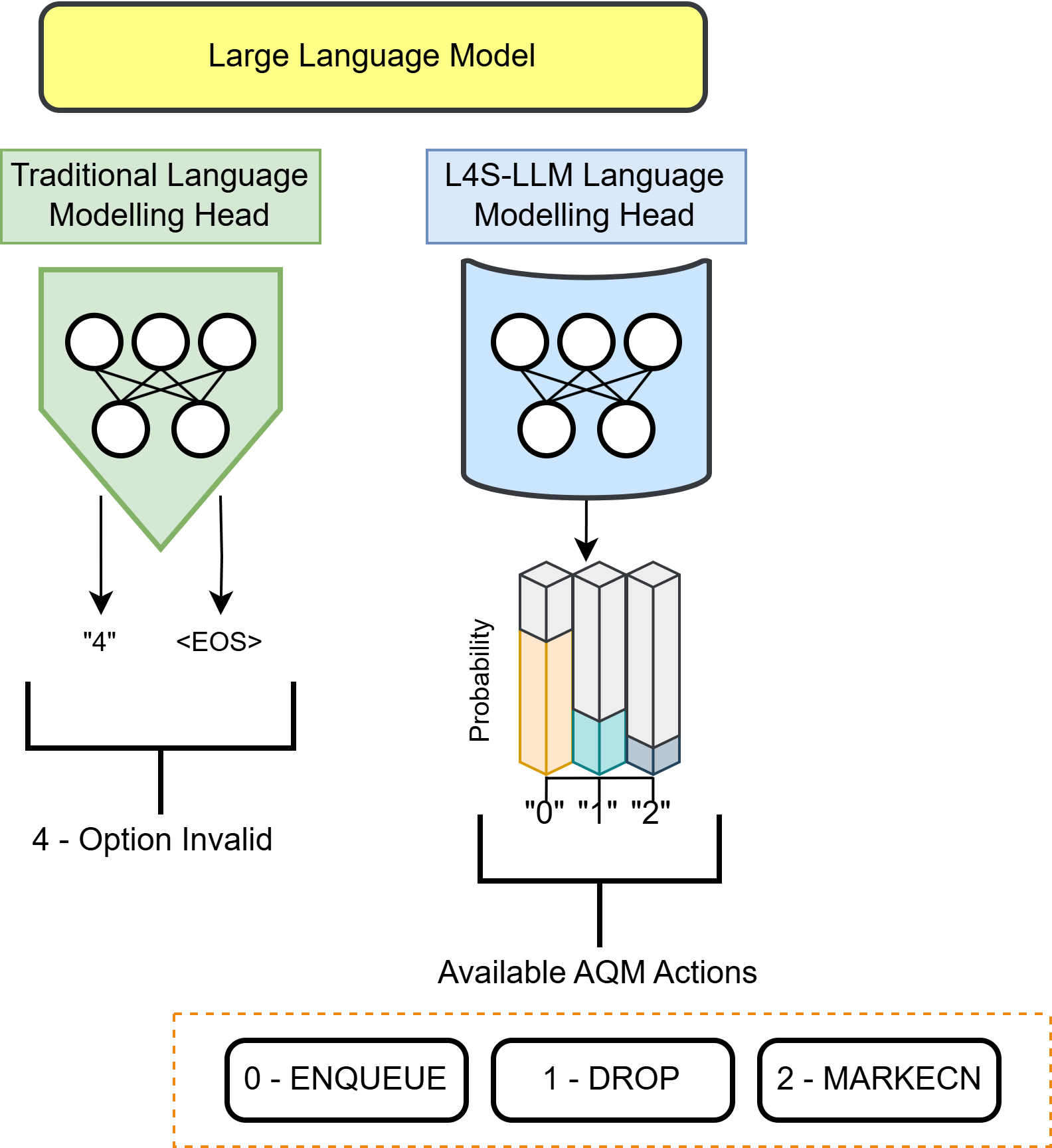}}
 \caption{L4S-LLM Head}
\label{fig:l4s_llm_headv2}
\end{figure}

Traditional language model heads generate responses token-by-token, requiring multiple inference rounds and introducing unacceptable latency for AQM applications. Our custom L4S-LLM head as shown in Figure~\ref{fig:l4s_llm_headv2} addresses this by generating complete actions in a single inference step. See details in Algorithm~\ref{algo:l4slm_head}.

\begin{algorithm}
\caption{L4S Language Modelling Head}
\label{algo:l4slm_head}
\begin{algorithmic}[1]
\State \textbf{Input:} $input\_embeddings, attention\_mask,$
\Statex $action\_embed\_pos, residual\_flag$
\State \textbf{Output:} predicted\_actions
\State \textbf{Initialize L4SLLM head:}
\Statex  \text{action\_head $\gets$ Linear($plm\_embed\_size, actionCount$)} \label{line:action_head}
\State action\_head $\gets$ action\_head.to(device) \Comment{Move to appropriate device}

\State logits\_used $\gets$ \texttt{Transformer}(input\_embeddings, attention\_mask) 
\If{residual\_flag}
    \State logits\_used $\gets$ logits\_used + input\_embeddings \Comment{Add residual connection}
\EndIf
\State \text{selected\_logits $\gets$ logits\_used[:, action\_embed\_pos - 2]} \Comment{Select relevant logits}
\State predicted\_actions $\gets$ action\_head(selected\_logits) \Comment{Predict actions using action head}
\State \textbf{Return} predicted\_actions \Comment{Probability Distribution of Candidate actions}
\end{algorithmic}
\end{algorithm}

The L4S-LLM head (line~\ref{line:action_head}, Algorithm~\ref{algo:l4slm_head}) is a trainable linear layer that maps transformer outputs directly to a probability distribution over the three possible actions: enqueue (0), drop (1), or ECN mark (2).\footnote{ This approach, inspired by speculative decoding, eliminates the need for token-by-token generation, reducing inference time from hundreds of milliseconds to under 40ms.}

\subsection{Data-Driven Reinforcement Learning (RL) Pipeline}
Recall that we employ RL-based adaptation pipeline which begins by collecting experience trajectories from the FreeBSD DualPI2 implementation. Each trajectory $\tau = \{r_t, s_t, a_t\}_{t=1}^{T}$ consists of states $s_t$ as 8-dimensional vectors of network metrics (including queue type, burst allowance, drop probability, current queue delay, accumulated probability, queue length in bytes, packet drops, and packet length), actions $a_t$ as discrete values (0: enqueue, 1: drop, 2: ECN mark), and rewards $r_t$ (for instance computed as $\text{pkt\_len}/(\text{qdelay}_t+1)$ in our implementation) to balance throughput and latency. For each trajectory, we evaluate the return $R_t = \sum_{i=t}^{T} \gamma^{i-t} r_i$ with discount factor $\gamma = 0.95$, representing cumulative future rewards from state $s_t$, transforming the trajectory into $\tau = \{R_t, s_t^1, \ldots, s_t^8, a_t\}_{t=1}^{T}$.

During training, we sample batches of data from the experience pool as $d = \{R_{t}, s_{t}^1, \ldots, s_{t}^8, a_{t}\}_{t=t-w+1}^{T}$ where $w$ is the context window, empirically determined to balance historical information against computational efficiency. The LLM processes this data to generate predicted actions $\{\hat{a}_{t}\}_{t=t-w+1}^{T}$, with training loss calculated as $L_{rl} = \frac{1}{w} \sum_{t=1}^{w} F_{rl}(a_{t}, \hat{a}_{t})$ where $F_{rl}$ is the cross-entropy loss function for our categorical action space. The model processes encoded states through the transformer architecture and generates action probabilities via the L4S-specific classification head, enabling the system to learn optimal congestion control policies from historical network behavior patterns.

\subsection{Low-Rank Adaptation (LoRA)}

\begin{figure}[t]
\centerline{\includegraphics[scale=0.456]{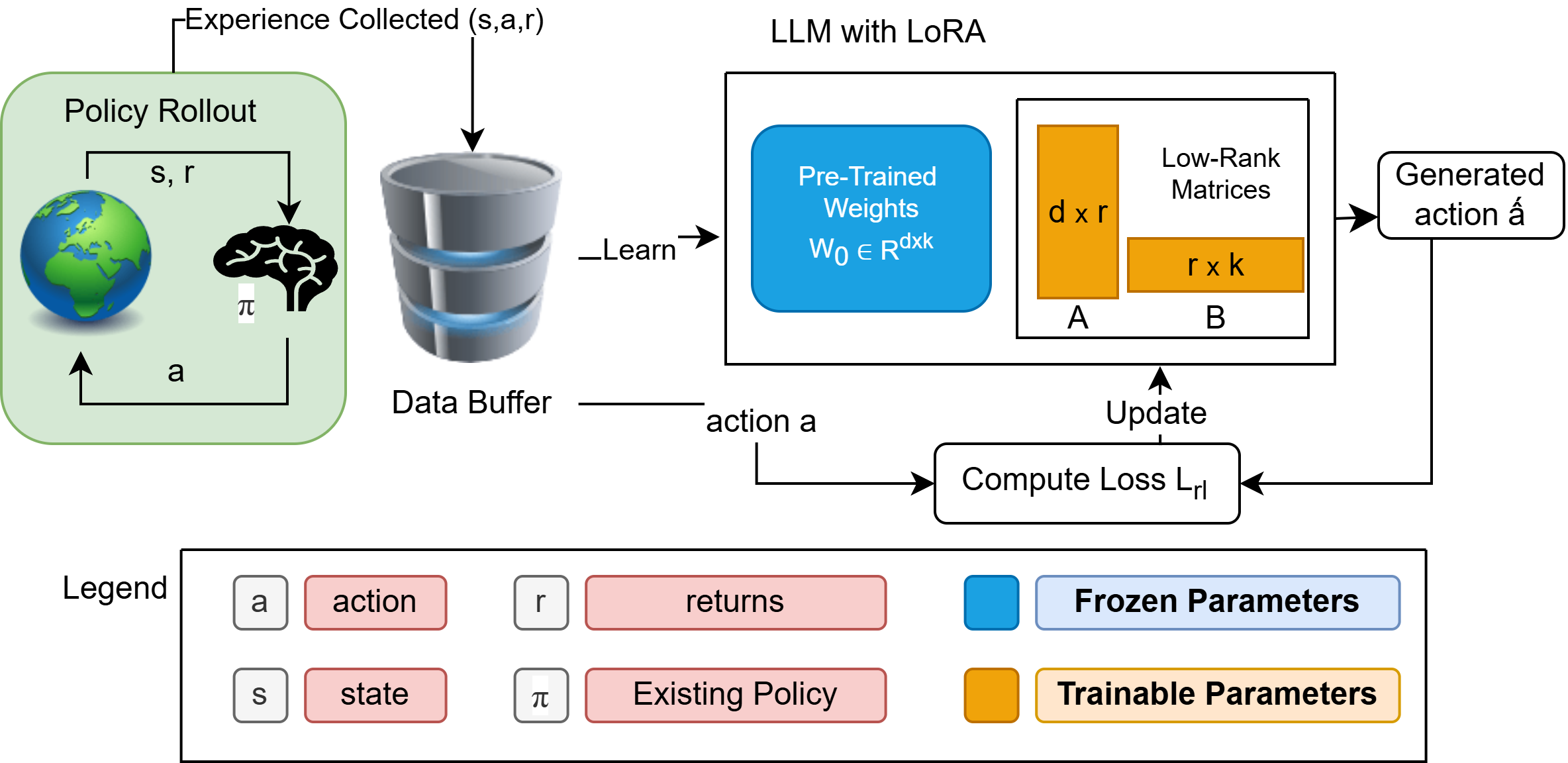}}
 \caption{Low-Rank L4S adaptation (LoRA) Design: Visualization of how LoRA matrices are integrated with frozen pre-trained weights to enable efficient fine-tuning.}
\label{fig:lora_diagram}
\end{figure}

Fine-tuning large language models with billions of parameters is computationally prohibitive for networking applications. We address this using Low-Rank Adaptation (LoRA), which dramatically reduces the number of trainable parameters while preserving model quality. Figure~\ref{fig:lora_diagram} shows the LoRA fine-tuning architecture where low-rank matrices (A and B) integrate with frozen pre-trained weights, enabling parameter-efficient L4S-LLM adaptation to network-specific patterns without full model retraining. Algorithm~\ref{algo:LoRA} details the LoRA training process. For each pre-trained weight matrix $W_0 \in \mathbb{R}^{d \times k}$ in the LLM, we introduce low-rank matrices $A \in \mathbb{R}^{d \times r}$ and $B \in \mathbb{R}^{r \times k}$ with rank $r \ll \min(d,k)$. The effective weight matrix 
\begin{equation}
W = W_0 + AB
\end{equation}
By freezing $W_0$ and only training $A$ and $B$, we reduce the number of trainable parameters from billions to millions—a 99\% reduction for Llama2-7B. 

\begin{algorithm}
\caption{Low-Rank Adaptation (LoRA) Algorithm \label{algo:LoRA}}
\begin{algorithmic}[1]
\State \textbf{Input:} Pretrained model $W_{\theta}$, adaptation rank $r$, training dataset $D$, number of epochs $E$
\State \textbf{Output:} Fine-Tuned model parameters $W_{fine\_tuned}$
\State Initialize low-rank matrices $A \in \mathbb{R}^{d \times r}$, $B \in \mathbb{R}^{r \times d}$ randomly
\State Set the learning rate $\eta$
\For{each epoch $e = 1$ to $E$}
    \For{each batch $b$ in the dataset $D$}
        \State Forward pass: compute model output with 
        \Statex \hspace{1.5cm} current $\theta$ and fine-tuned weights $(A, B)$
        \State Compute the loss $\mathcal{L}$ based on the model output 
        \Statex \hspace{1.5cm} and ground truth
        \State Compute the gradient of the loss with respect to $\theta$ 
        \Statex \hspace{1.5cm} and $(A, B)$
        \State Update the low-rank matrices using gradient descent:
        \[
        A \gets A - \eta \frac{\partial \mathcal{L}}{\partial A}, \quad B \gets B - \eta \frac{\partial \mathcal{L}}{\partial B}
        \]
    \EndFor
\EndFor
\State Set the final Fine-Tuned model parameters:
\[
W_{fine\_tuned} \gets W_{\theta} + A B
\]
\State \textbf{Return:} $\theta_{adapt}$
\end{algorithmic}
\end{algorithm}

For Llama2-7B, we use rank $r=128$, reducing trainable parameters from 7 billion to approximately 70 million while maintaining 97.56\% accuracy. This approach preserves the pre-trained knowledge in the LLM while enabling efficient adaptation to the AQM domain.

\color{black}

\section{Setup and Evaluation}
\label{aqmeval}

\subsection{Experimental Testbed Setup}

\begin{figure}[htbp]
\centerline{\includegraphics[width=0.79\linewidth]{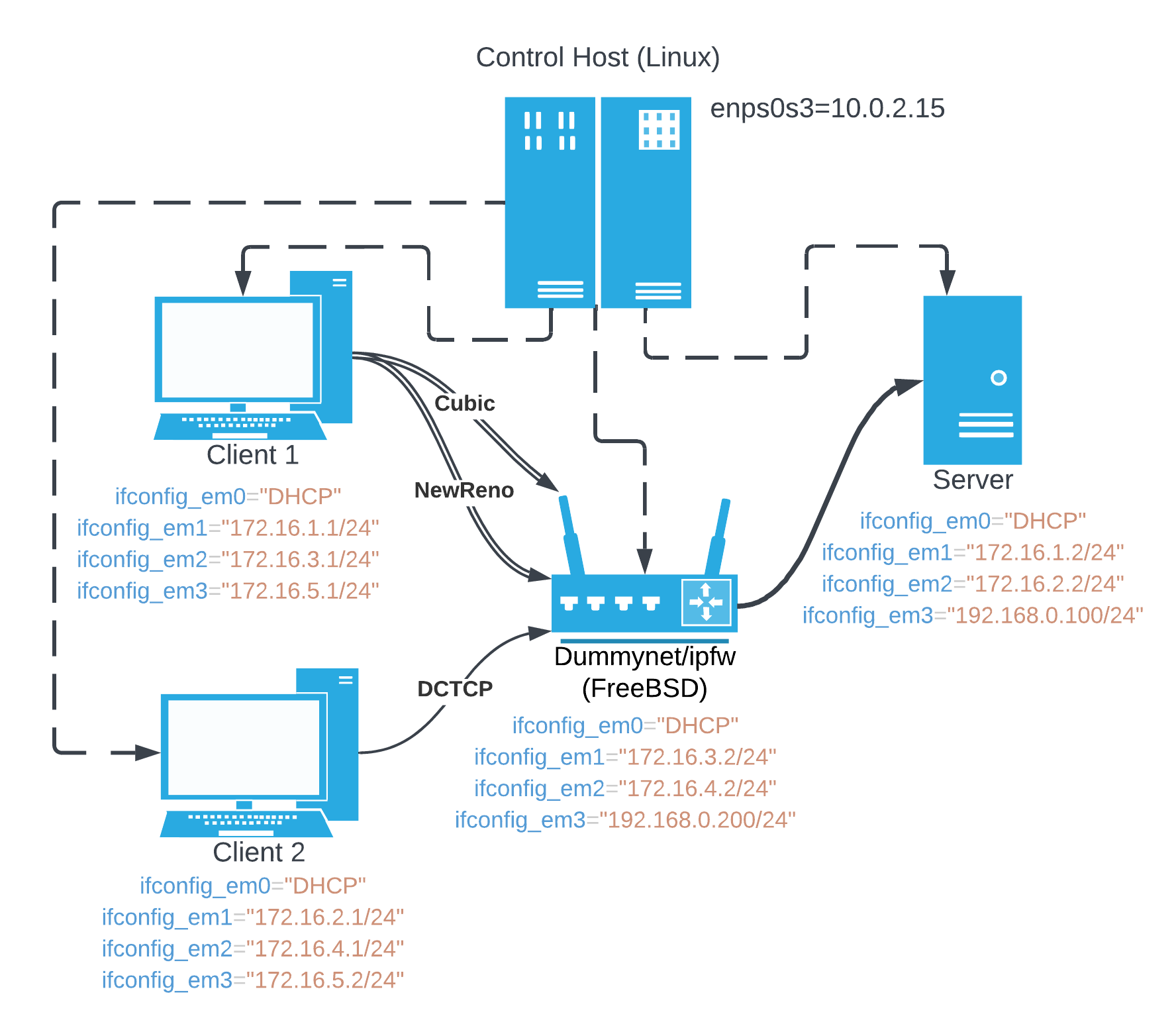}}
 \caption{Topology of the developed experimental testbed for collecting datasets and evaluating L4S-LLM framework.}
\label{fig:networktopologya}
\end{figure}
\color{black}
We collect the experiences required for training our L4S-LLM by developing a testbed as shown in Figure~\ref{fig:networktopologya} consisting of FreeBSD-based virtual machines, built and configured using VirtualBox and VMware virtualization platforms. The virtual machines run FreeBSD version 14.1 as their operating system. To create a network setup in VirtualBox, we clone four virtual machines: two clients (Client 1 and Client 2), one router, and one server. Client 1 utilizes two different congestion control algorithms - Cubic and NewReno simultaneously, while Client 2 uses ECT(1)-enabled DCTCP congestion control algorithm. We replace DCTCP with UDP Prague and repeat all the experiments in VMware.\footnote{VirtualBox provides open-source accessibility, simpler network configuration options, better FreeBSD guest support through built-in additions, and lower resource requirements suitable for basic protocol testing. In contrast, VMware delivers superior network performance with enhanced throughput and lower latency, advanced virtual networking capabilities (distributed switches, VLAN trunking, traffic shaping), better hardware virtualization, and more reliable snapshot functionality for experimental state preservation. For performance-sensitive experiments like congestion control or AQM evaluation, VMware's network stack more accurately emulates physical hardware behavior, while VirtualBox remains adequate for functional protocol testing with reduced resource overhead.} The router is configured with a custom-built DualPI2 algorithm that supports the L4S architecture, as described in \cite{freebsdl4s}. All clients send data to the server through the router as illustrated in Figure~\ref{fig:networktopologya}. All scripts and additional information required to set up the environment are made public in the \href{https://github.com/MPTCP-FreeBSD/L4S-LLM.git}{L4S-LLM repository}.

In our experiment to collect data, we configure the router with a bandwidth of 8 Mbps and a delay of 10 ms. We initiate two \texttt{iperf3} connections for each congestion control algorithm used in Client 1, with each connection utilizing two parallel flows via the \texttt{-P 2} option in \texttt{iperf3}. Client 2 only utilizes a single flow for its \texttt{iperf3} connection. Each \texttt{iperf3} test runs for one minute. We repeat the experiment ten times and collect all the data. We have enabled ECN on all devices.

To gather the data, we use a debugging tool that is directly embedded in the DualPI2 kernel code to store data from the kernel into the debug kernel logs. The relevant data is then extracted from these logs. If the size of the collected data exceeds 1000 KB, the log files will be truncated and rotated to prevent excessive growth. To increase the file size limits, we modify the configuration files for the \texttt{newsyslog} service, found in \textit{/etc/newsyslog.conf}.

Once the data collection process is complete, we process the collected data into an experience pool to be fed into the LLM; see \href{https://github.com/MPTCP-FreeBSD/LLM\_Gen\_Exp\_Pool}{L4S-LLM experience pool}\footnote{https://github.com/MPTCP-FreeBSD/LLM\_Gen\_Exp\_Pool} implementation for details. We experiment with multiple foundation LLM models, including Llama2\footnote{https://www.llama.com/llama2}\cite{touvron2023llama} (7b), OPT (1.3b), GPT2\footnote{https://openai.com/index/better-language-models/}\cite{openai2024chatgpt} (125m) and T5-LLM\footnote{https://github.com/google-research/text-to-text-transfer-transformer}\cite{T5048643}, which are fine-tuned for our L4S-LLM task. Among these, Llama2 demonstrated the best performance (see Section III-B) and was selected as the foundational LLM for extensive evaluation in simulations, where L4S-LLM is compared against the legacy AQM algorithms. To execute L4S-LLM, we utilize \texttt{Runpod}, an online cloud computing platform that provides access to powerful GPUs and scalable storage, ensuring efficient training and testing of our framework. We used two NVIDIA A40 GPUs, each equipped with 48 GB of VRAM, carried out on an Ubuntu 22.04 instance with CUDA 11.8 version.

We log key metrics during L4S-LLM training and testing, including loss, accuracy, reward, GPU/CPU utilization, computation time, RAM usage, and total training/testing time. To evaluate the model's efficiency, we analyze loss and accuracy, which are calculated using cross-entropy loss for this multi-class classification problem. Cross-entropy loss measures the discrepancy between predicted probabilities and true labels, outputting a scalar value where lower values indicate better performance. The training process employs mini-batches with gradient accumulation to enhance efficiency and stability. Gradient clipping is utilized to prevent gradient explosion by capping gradients to a predefined threshold during backpropagation. The model predicts a probability distribution over actions, represented by \texttt{action.preds}, and the predicted action is obtained using the \texttt{argmax()} function. True actions, extracted and normalized from the experience pool, are aligned with the predicted actions using a custom vectorization and conversion function to ensure compatibility for computing accuracy\begin{align}
\text{Accuracy}= \frac{1}{N} \sum_{i=1}^{N} \mathbb{I}(\text{preds\_actions}_i = \text{true\_classes}_i)
\end{align}
over all N samples or steps in the epoch where \( \mathbb{I} \) is the indicator function returns 1 when the predicted action (preds\_actions) matches the true action class (true\_classes), and 0 otherwise.

To ensure robust generalization beyond training scenarios, we implemented multiple strategies to counteract limitations of supervised distillation. We systematically varied key network parameters (bandwidth, RTT, queue sizes, traffic patterns) during data collection to prevent model memorization. Our approach incorporated queue perturbations, temporal shifting of metrics, and diverse traffic traces from multiple congestion control algorithms to create novel training examples. 
\color{black}

\subsection{L4S-LLM Training and Validation}
\subsubsection{Evaluation of Loss Dynamics in L4S-LLM}

\begin{figure}[h!]
\centerline{\includegraphics[width=0.67\linewidth]{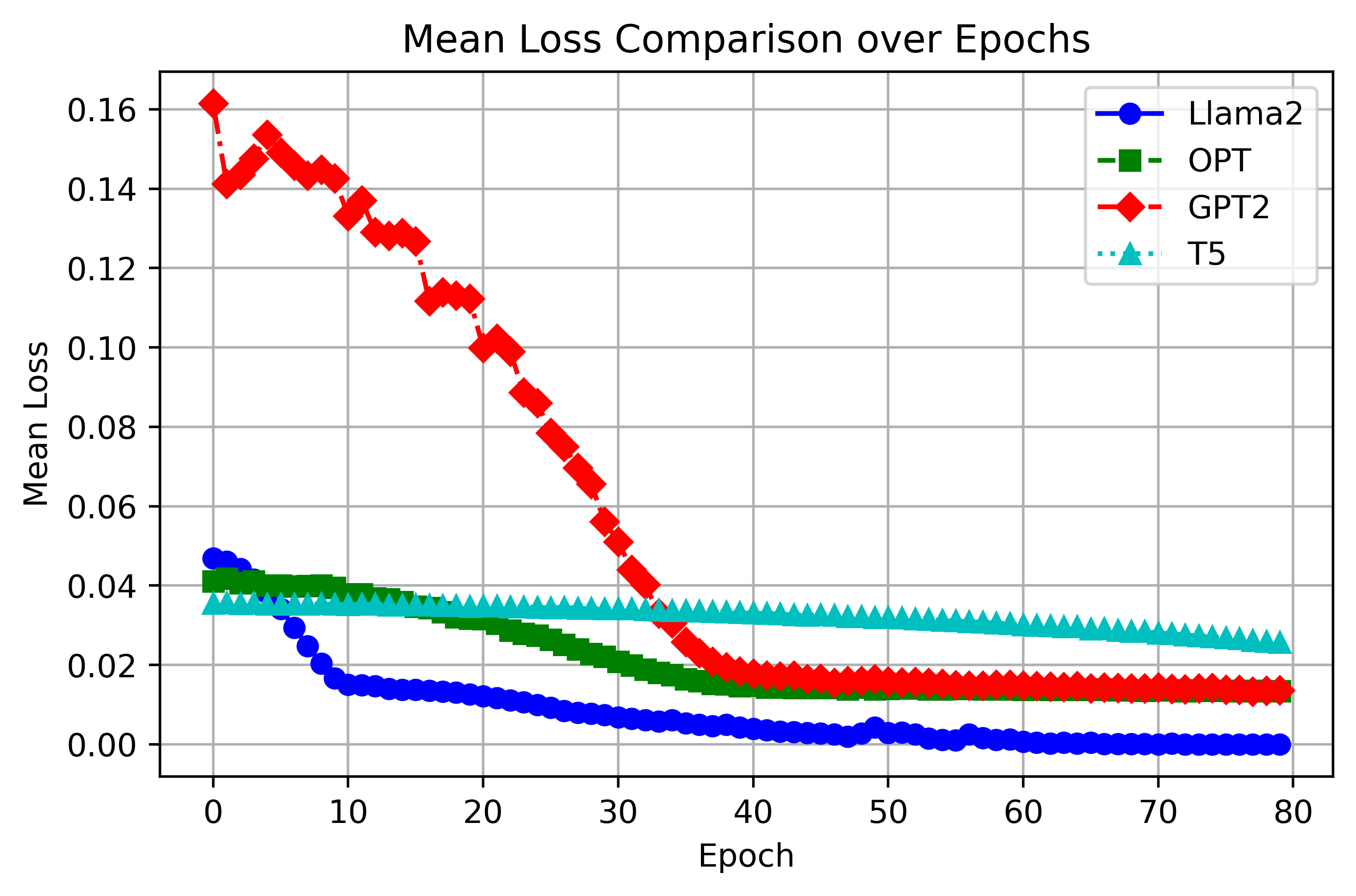}}
 \caption{L4S-LLM Training Loss Trends: Evolution of Mean Loss over 80 Epochs }
\label{fig:trainingloss}
\end{figure}

%
\color{black}
Figure~\ref{fig:trainingloss} demonstrates the training loss evolution across four LLM architectures over 80 epochs. Llama2 exhibits superior convergence characteristics, achieving the lowest mean loss by epoch 5 and maintaining consistent improvement throughout training. OPT displays gradual loss reduction with slower optimization dynamics, while T5-LLM follows similar trends but achieves marginally better final loss than OPT. GPT2 initially exhibits the highest loss but rapidly converges from epoch 40, ultimately matching OPT's performance level.

The loss trajectories validate L4S-LLM's robust adaptability to complex network queue management across diverse architectural foundations. By leveraging real-time kernel-level metrics, the framework demonstrates effective optimization and generalization under dynamic network conditions. Llama2's 7B parameter architecture achieves optimal loss minimization, followed by OPT and GPT2 with comparable performance, while T5-LLM exhibits slightly elevated loss values. These results reinforce Llama2's selection as the primary foundation model, combining superior convergence properties with effective parameter reduction through LoRA for practical deployment in network congestion control scenarios.   
\color{black}



\subsubsection{Evaluation of Accuracy Dynamics in L4S-LLM}

\begin{figure}[htbp]
\centerline{\includegraphics[width=0.67\linewidth]{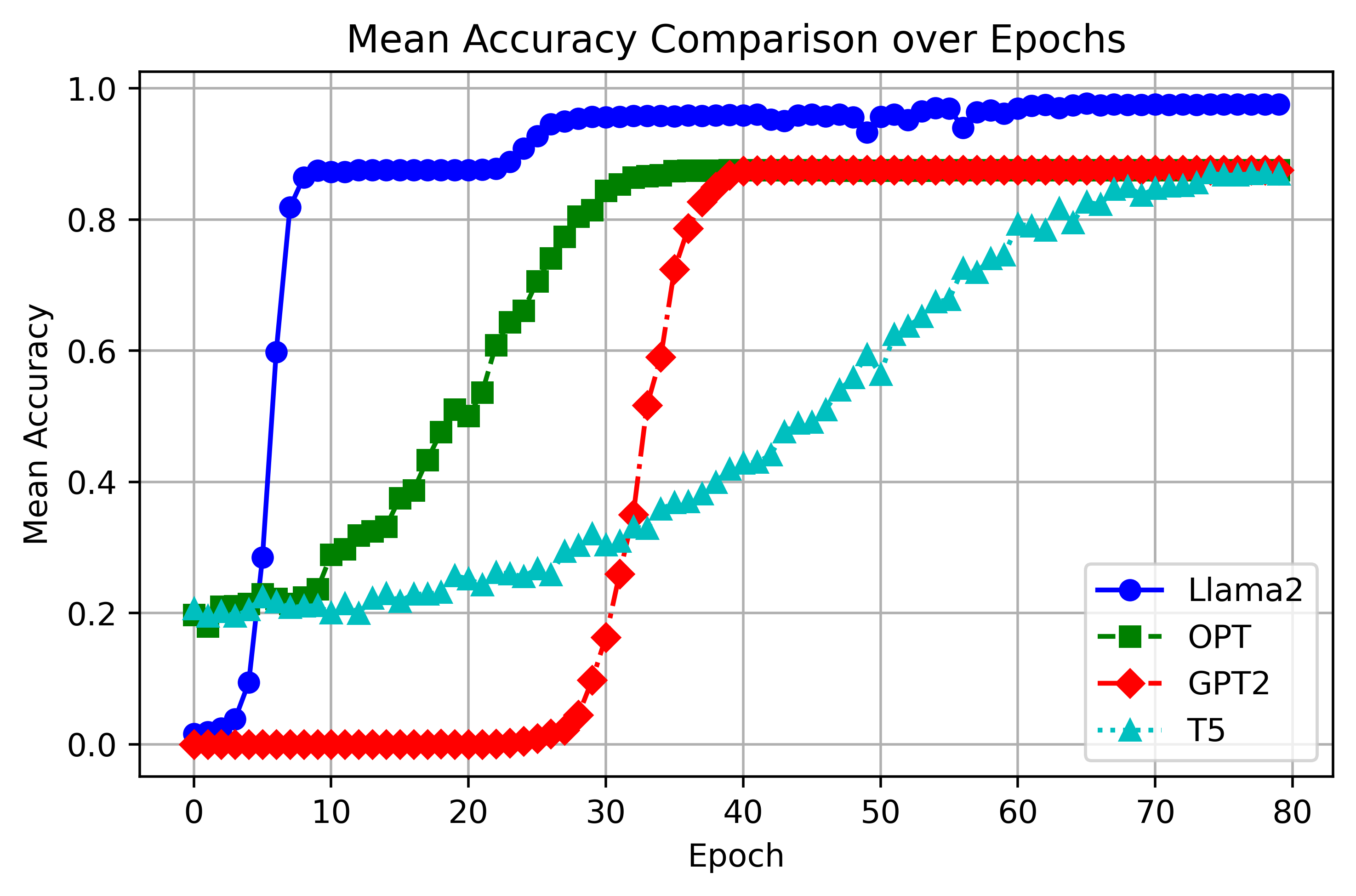}}
 \caption{L4S-LLM Training Accuracy Trends: Mean Accuracy over 80 Epochs }
\label{fig:trainingaccuracy}
\end{figure}

\color{black}
Figure~\ref{fig:trainingaccuracy} illustrates the training accuracy evolution across four LLM architectures during fine-tuning. Llama2 exhibits initial near-zero accuracy with rapid improvement between epochs 5-9, followed by plateau and gradual ascent from epoch 30, ultimately achieving superior performance at 97.56\%. OPT demonstrates higher initial accuracy than competitors, steadily increasing until epoch 40 where it stabilizes at 83\%. GPT2 remains stagnant near zero until epoch 29, then exhibits sharp improvement through epoch 40 before stabilization. T5-LLM begins at 20\% accuracy with gradual progression, plateauing around 83\% after epoch 75.

Llama2 significantly outperforms all alternatives, achieving 97.56\% accuracy compared to the 82-83\% plateau reached by OPT, GPT2, and T5-LLM. Despite requiring over 20 hours training time versus competitors' sub-hour requirements, Llama2's superior accuracy justifies the computational investment. Critically, LoRA reduces Llama2's 7B parameters to 70M trainable parameters while maintaining performance, and inference times remain comparable to smaller models during deployment. 
\color{black}



\subsection{Evaluation of L4S-LLM over AQM}

We evaluate the performance of our L4S-LLM framework, with Llama2 -7B as our base, fine-tuned using LoRA, with its performance benchmarked against real-world L4S AQM data. The benchmarking focuses primarily on optimizing queue delay and bandwidth utilization. The network topology depicted in Figure~\ref{fig:networktopologya} involves Client 1 using Cubic and New-Reno concurrently, while Client 2 is running DCTCP, and the router is configured to implement L4S. Using real-world data LLM predicts the AQM action - Enqueue, Dequeue, or ECN marking. The LLM is used at specific intervals as it is resource-intensive. Our results, Figs.~\ref{fig:qdboxComparison} and~\ref{fig:bpqdboxComparison}, demonstrate the LLM's ability to optimize the queue delay and bandwidth of both DCTCP and UDP Prague flows.

\subsubsection{Analysis of Queue Delay}

\begin{figure}[!t]
    \centering
    \subfloat[Aggregate Delay with 10-Interval LLM Inputs with DCTCP]{%
        \includegraphics[width=0.45\linewidth]{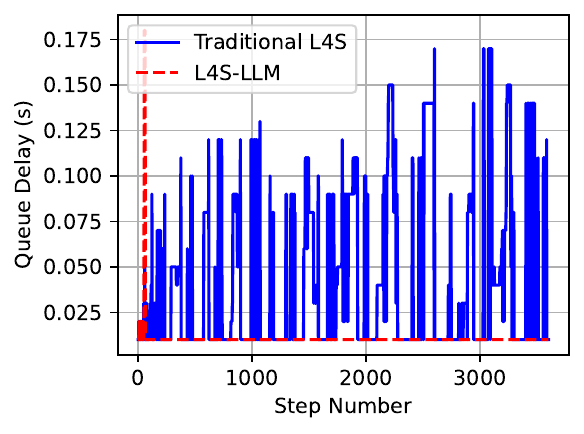}%
        \label{fig:qdbox100}
    }
    \hfill
    \subfloat[Aggregate Delay with 100-Interval LLM Inputs with DCTCP]{%
        \includegraphics[width=0.45\linewidth]{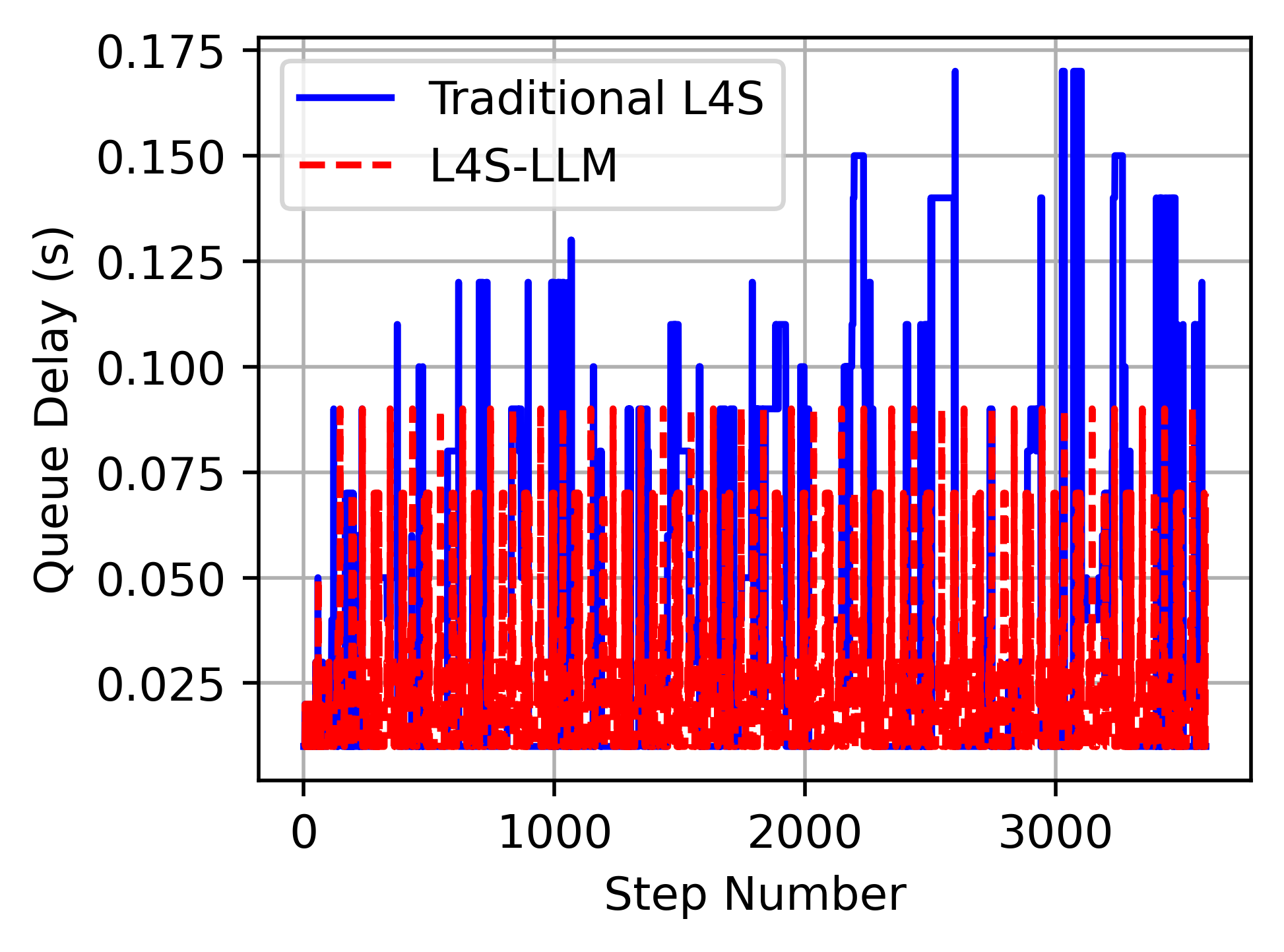}%
        \label{fig:qdbox10}
    } \vfill
      \subfloat[Classic Queue with 10-Interval LLM Inputs with UDP Prague ]{%
        \includegraphics[width=0.45\linewidth]{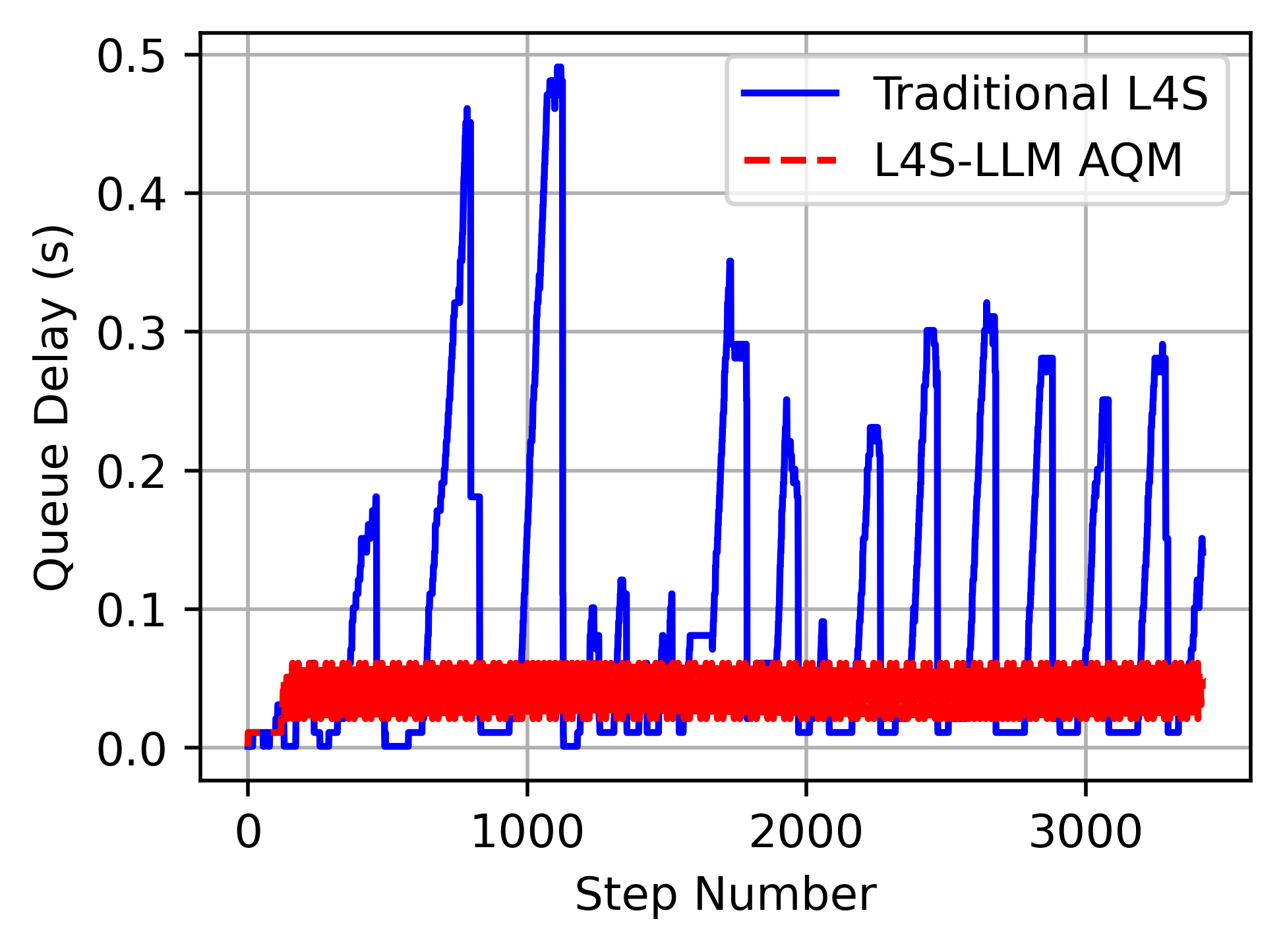}%
        \label{fig:qdbox100}
    }
    \hfill
    \subfloat[Classic Queue with 100-Interval LLM Inputs with UDP Prague]{%
        \includegraphics[width=0.45\linewidth]{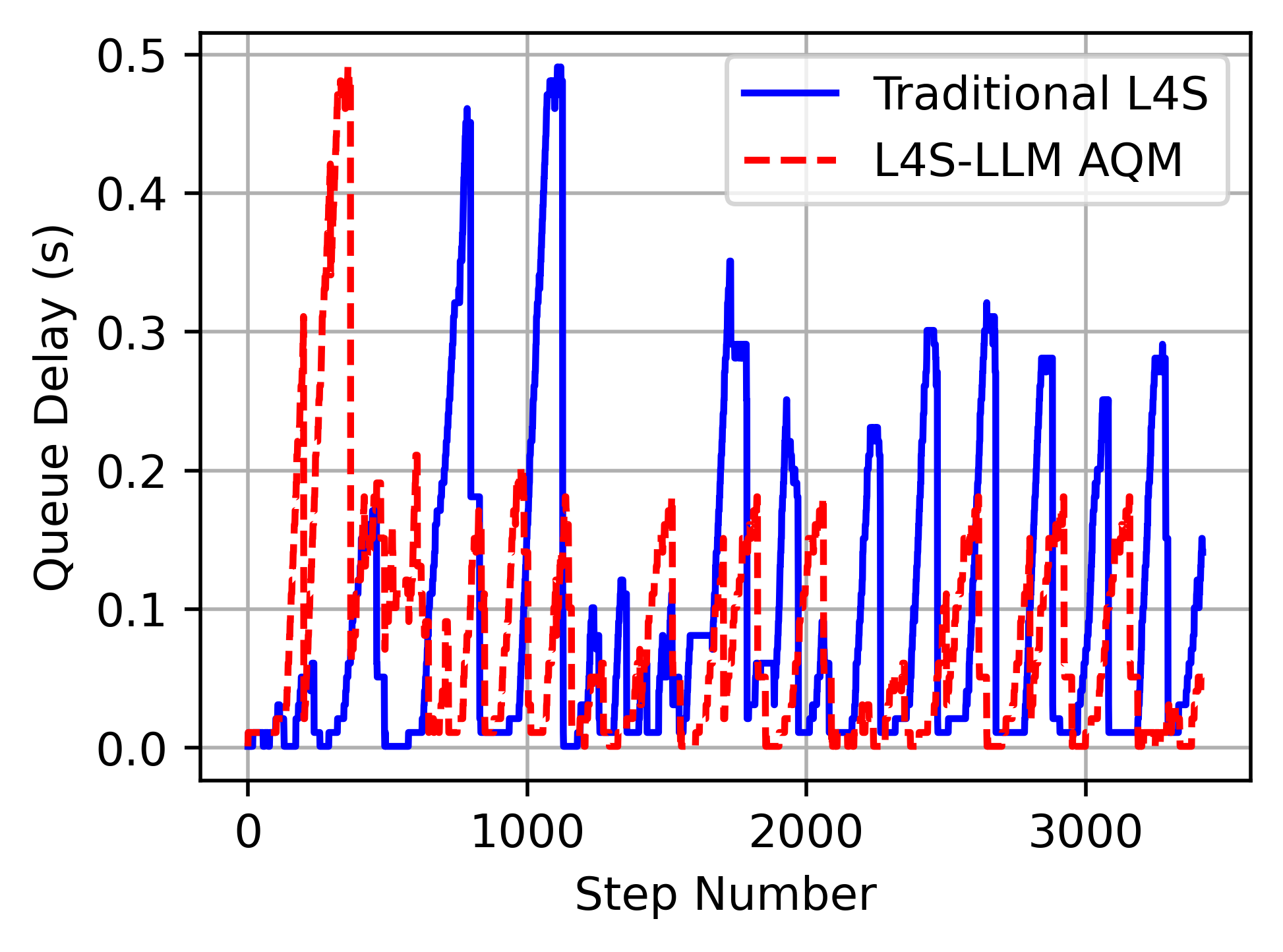}%
        \label{fig:qdbox10}
    }
    \caption{Queue Delay Analysis with Periodic LLM Inputs}
    \label{fig:qdboxComparison}
\end{figure}

\begin{figure}[!t]
    \centering
    \subfloat[DCTCP: 10-Interval Inputs]{%
        \includegraphics[width=0.45\linewidth]{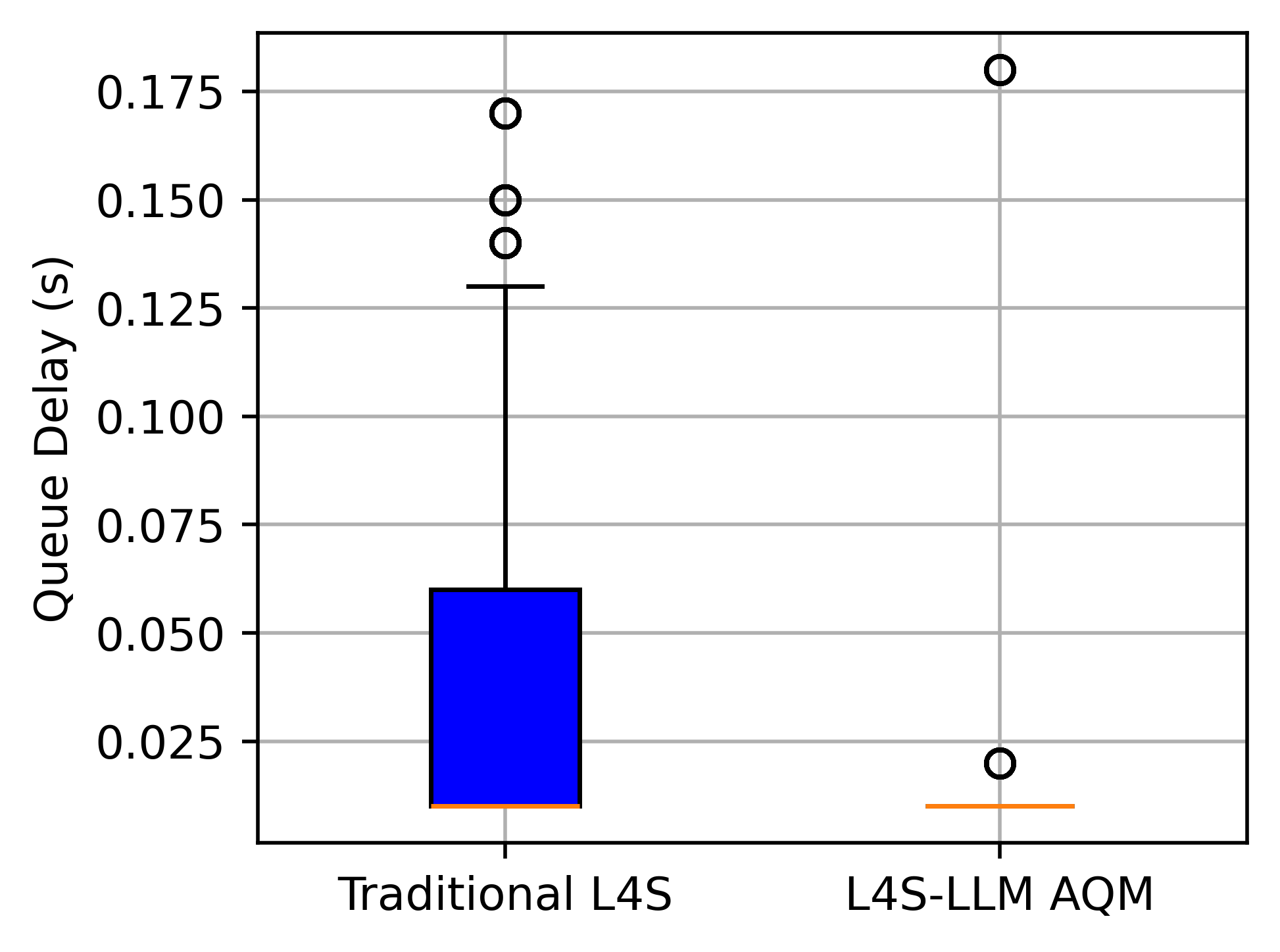}%
        \label{fig:bpqdbox100}
    }
    \hfill
    \subfloat[DCTCP: 100-Interval Inputs]{%
        \includegraphics[width=0.45\linewidth]{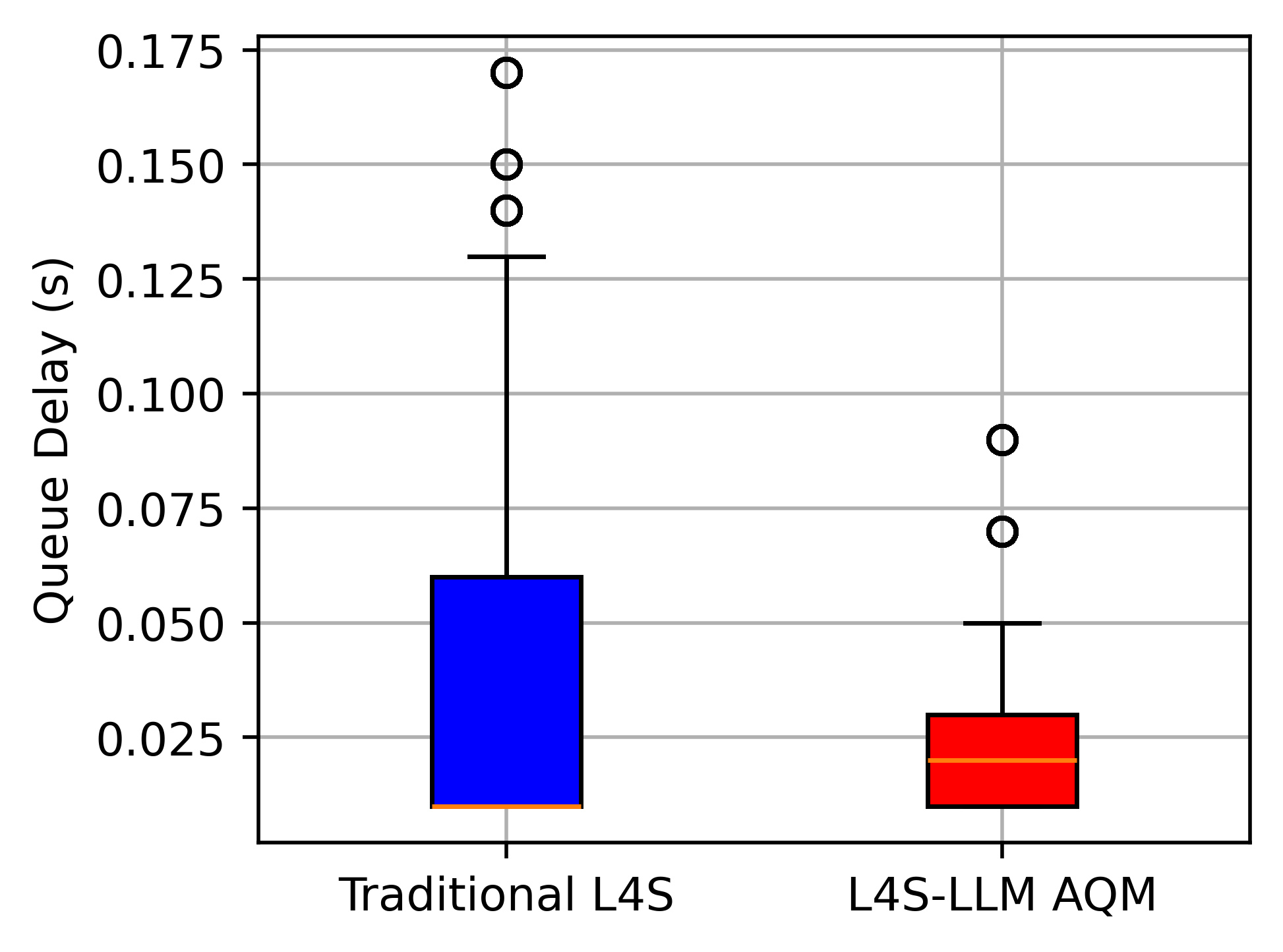}%
        \label{fig:bpqdbox10}
    }\vfill
       \subfloat[UDP Prague: 10-Interval Inputs]{%
        \includegraphics[width=0.45\linewidth]{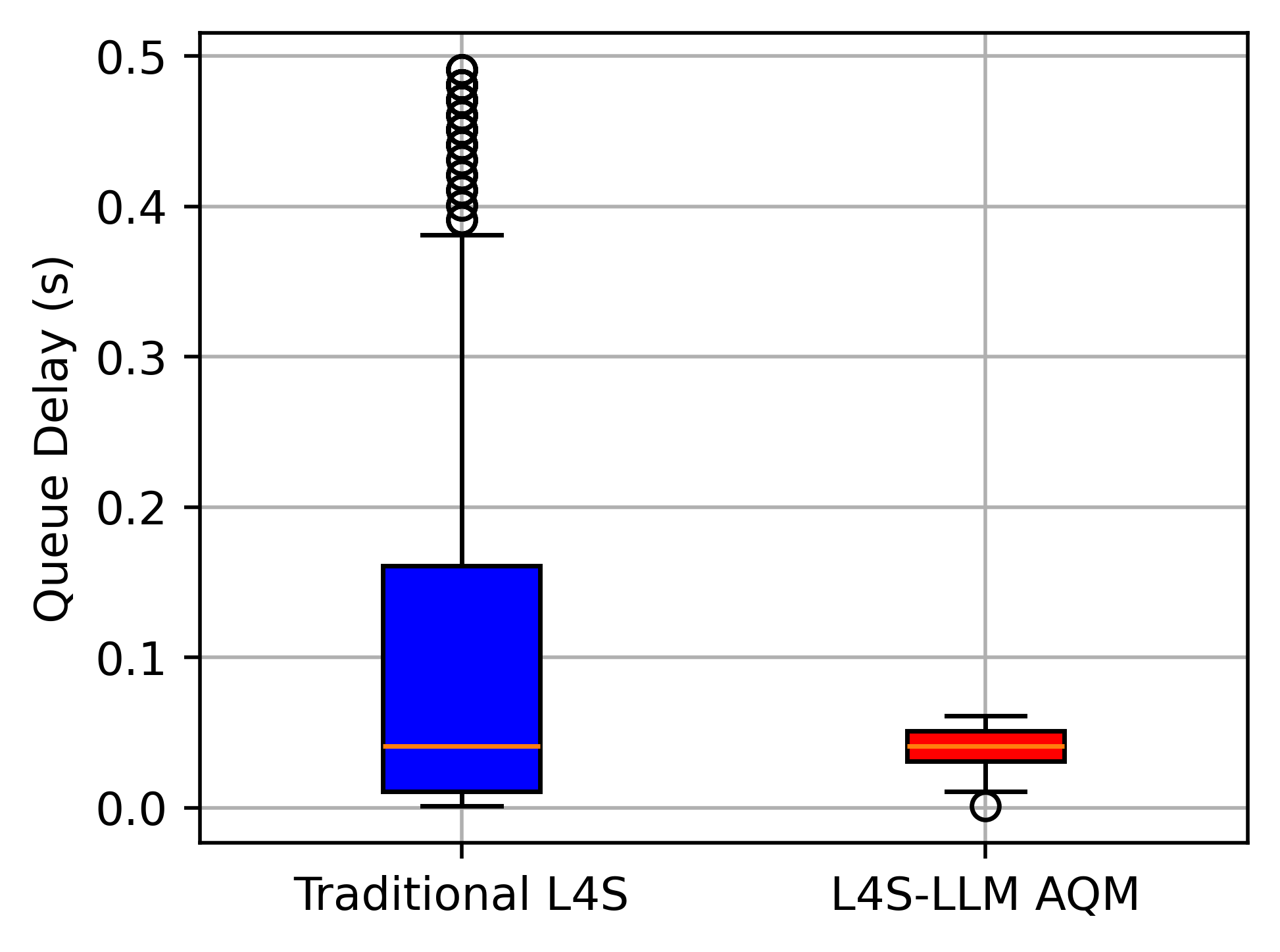}%
        \label{fig:bpqdbox100}
    }
    \hfill
    \subfloat[UDP Prague: 100-Interval Inputs]{%
        \includegraphics[width=0.45\linewidth]{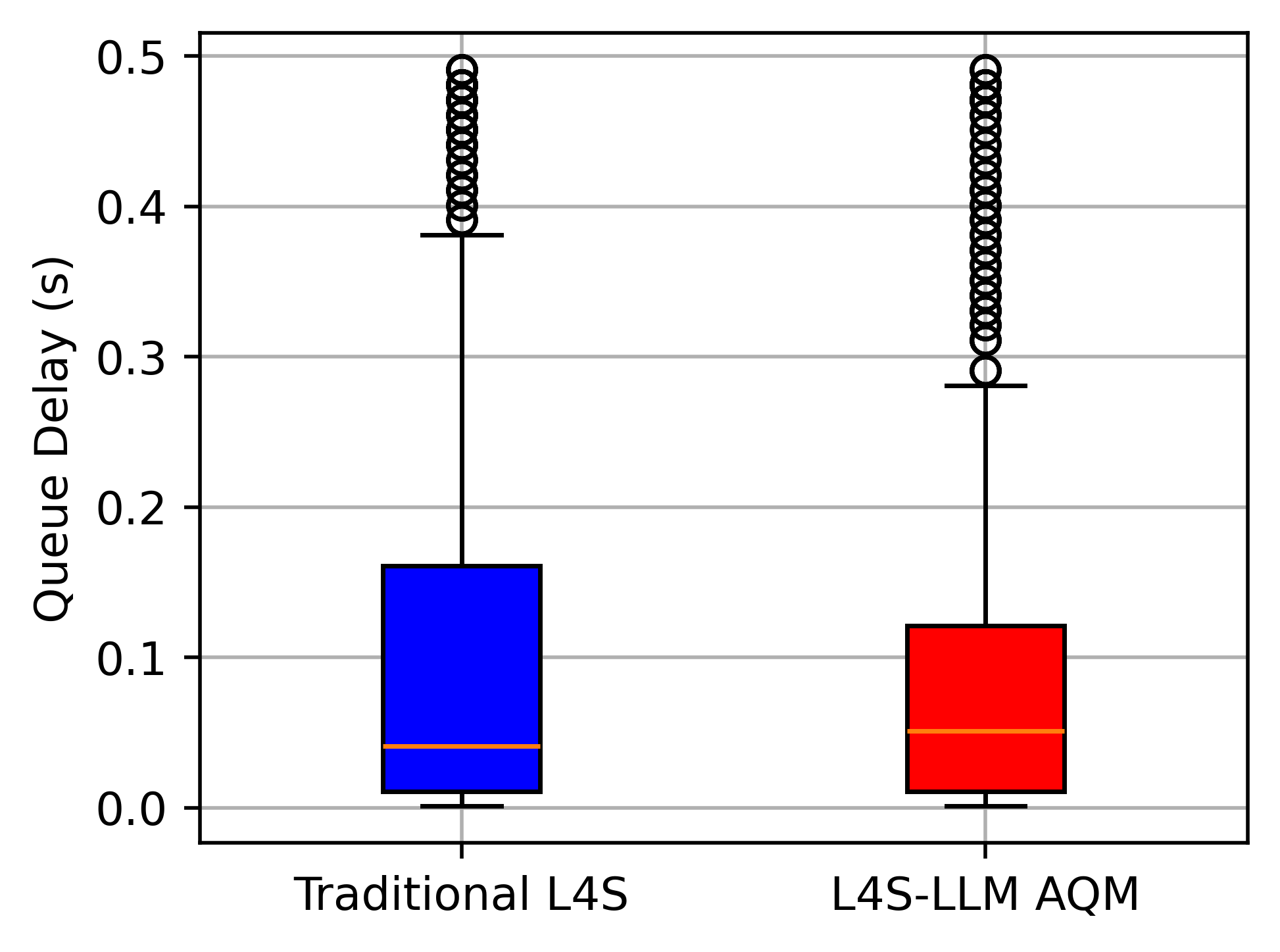}%
        \label{fig:bpqdbox10}
    }
    \caption{Box Plot of DCTCP and UDP Prague Queue Delays of Fig 11.}
    \label{fig:bpqdboxComparison}
\end{figure}

Figure~\ref{fig:qdboxComparison} presents a comparison of queue delays between the original L4S AQM implementation and the enhanced L4S-LLM framework leveraging the fine-tuned LLM model to optimize queue delay. The LLM model determines the most optimal action -- Enqueue, Drop and ECN marking -- based on the system's state. As shown in Figure~\ref{fig:bpqdboxComparison}, our reward function guides the  LLM in finding the most optimal action which keeps improving the reward. The results show a significant reduction in queue delay for the LLM-based system compared to the traditional algorithm. The traditional queue delay has a higher median queue delay with high variability. Conversely, the L4S-LLM achieves a remarkably low median queue delay value, with low variability and fewer outliers, demonstrating the LLM's capacity to maintain low-latency behaviour. Our extensive analysis hereafter primarily focuses on DCTCP/VirtualBox, though results for UDP Prague/VMware have been produced using the code available in our GitHub repository. 


\begin{figure}[!t]
    \centering
    \subfloat[CDF of Queue Delay with 10-Interval LLM Inputs]{%
        \includegraphics[width=0.45\linewidth]{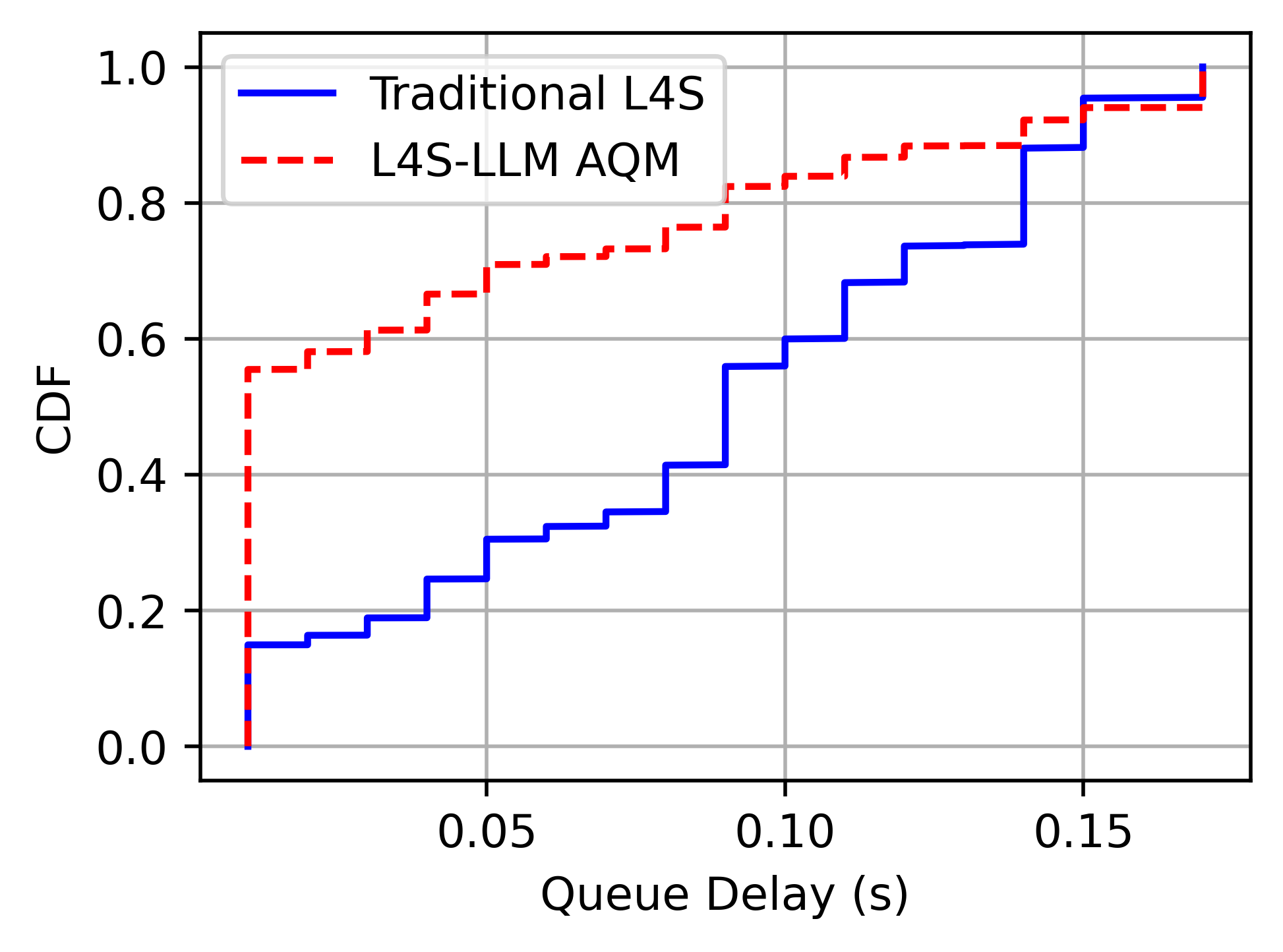}%
        \label{fig:qdcdf10}
    }
    \hfill
    \subfloat[CDF of Queue Delay with 100-Interval LLM Inputs]{%
        \includegraphics[width=0.45\linewidth]{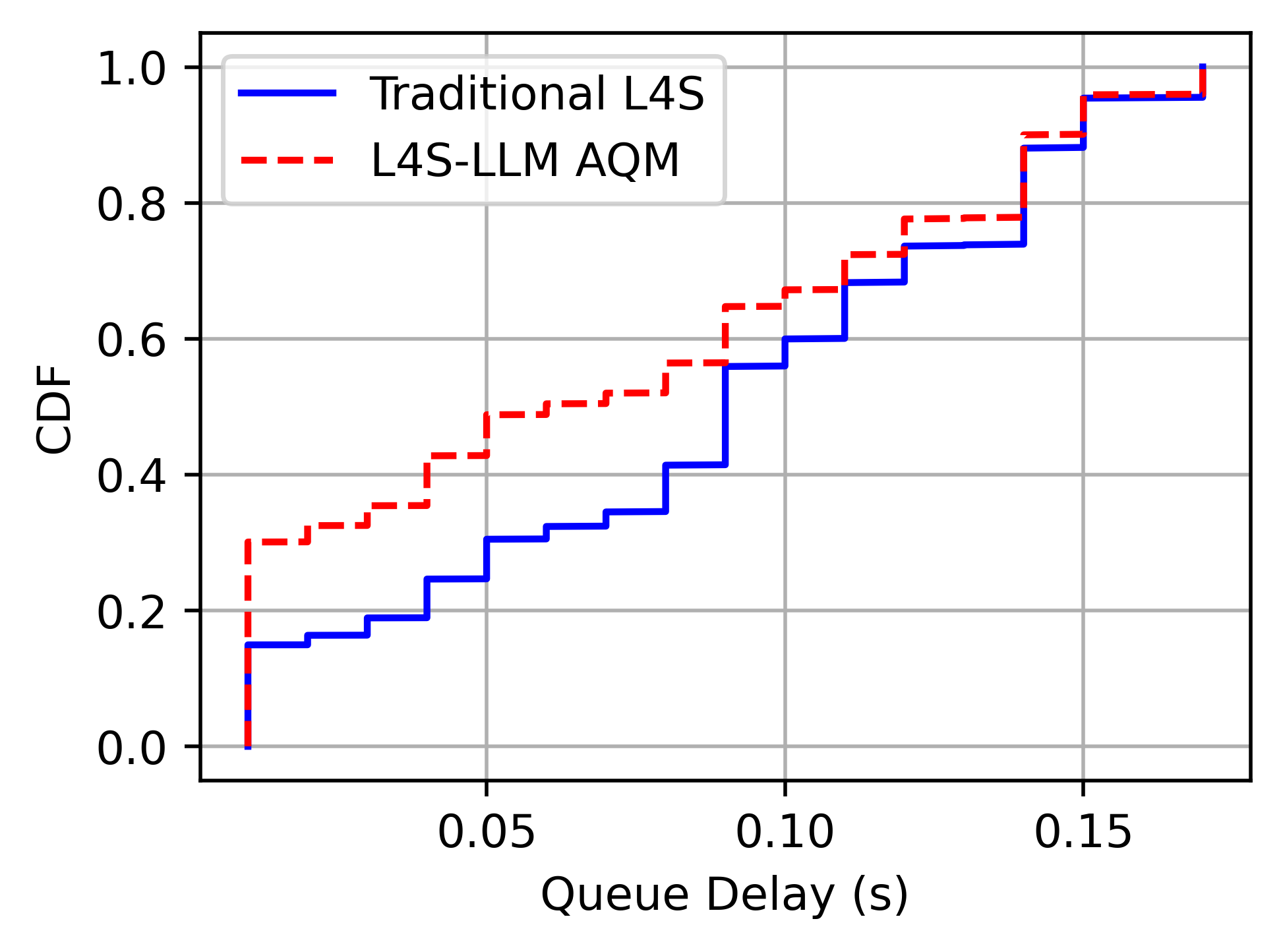}%
        \label{fig:qdcdf100}
    }
    \caption{CDF of DCTCP Queue Delay with Periodic LLM Inputs.}
    \label{fig:qdcdfComparison}
\end{figure}

Figure~\ref{fig:qdcdfComparison} illustrates the cumulative distribution functions (CDFs) of queue delay for traditional L4S and L4S-LLM AQM. These results provide an alternative representation of queue delay performance dynamics capturing LLM's effectiveness in optimizing queue management strategies.

\subsubsection{Analysis of Bandwidth utilization}


\begin{figure}[h!]
    \centering
    \subfloat[Bandwidth utilization with 10-Interval LLM Inputs]{%
        \includegraphics[width=0.45\linewidth]{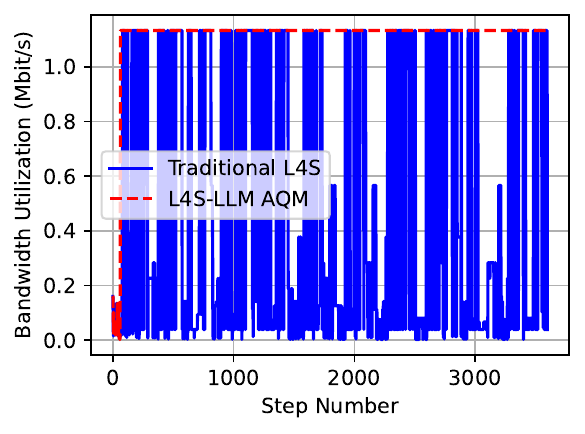}%
        \label{fig:thrptbox10}
    }
    \hfill
    \subfloat[Bandwidth utilization with 100-Interval LLM Inputs]{%
        \includegraphics[width=0.45\linewidth]{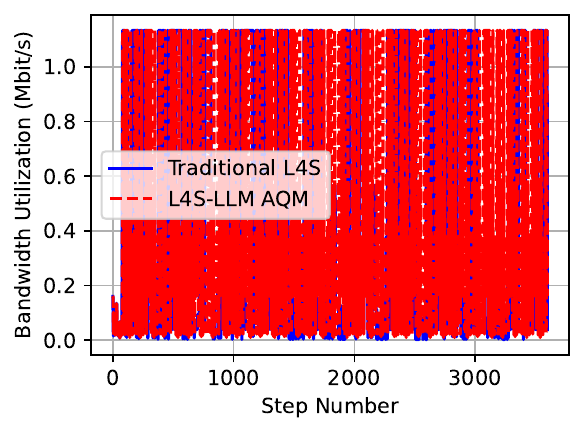}%
        \label{fig:thrptbox100}
    }
    \caption{Comparision of Bandwidth utilization when L4S-LLM is enabled at 10 and 100-Packet Intervals.}
    \label{fig:thrptboxComparison}
\end{figure}

\begin{figure}[h!]
    \centering
    \subfloat[Box Plot of Bandwidth utilization with 10-Interval LLM Inputs]{%
        \includegraphics[width=0.45\linewidth]{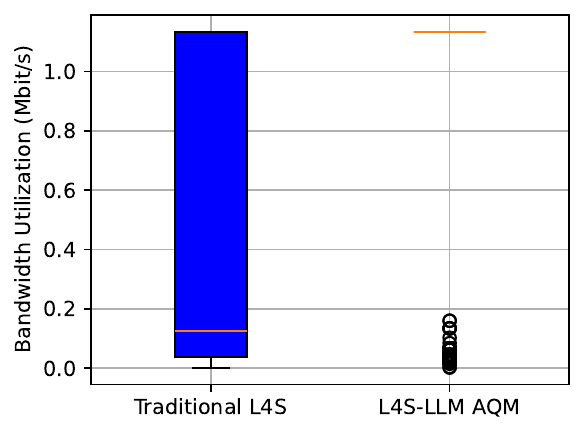}%
        \label{fig:bpthrptbox10}
    }
    \hfill
    \subfloat[Box Plot of Bandwidth utilization with 100-Interval LLM Inputs]{%
        \includegraphics[width=0.45\linewidth]{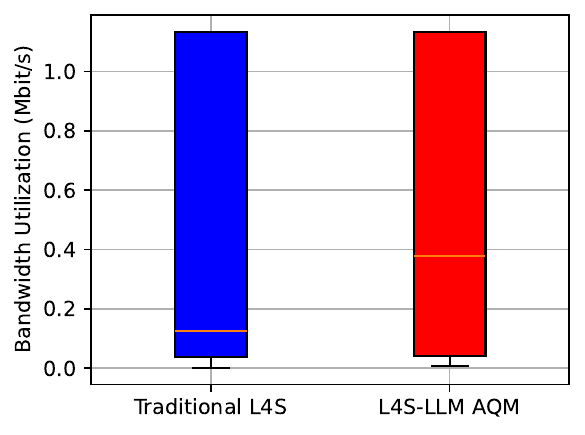}%
        \label{fig:bpthrptbox100}
    }
    \caption{Box Plot of Bandwidth utilization with LLM Enabled at 10 and 100-Packet Intervals.}
    \label{fig:bpthrptboxComparison}
\end{figure}

\begin{figure}[htbp]
    \centering
    \subfloat[CDF of Bandwidth utilization with 10-Interval LLM Inputs]{%
        \includegraphics[width=0.45\linewidth]{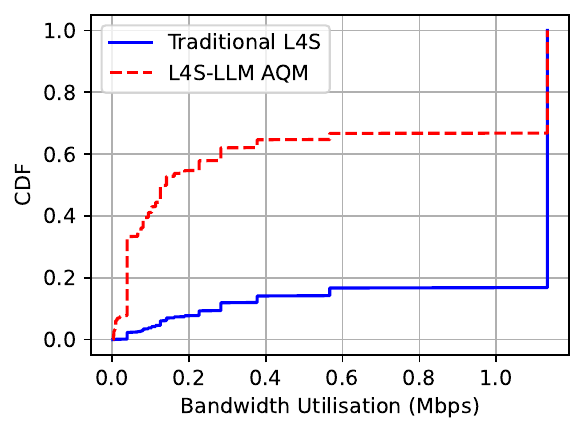}%
        \label{fig:thrptcdf10}
    }
    \hfill
    \subfloat[CDF of Bandwidth utilization with 100-Interval LLM Inputs]{%
        \includegraphics[width=0.45\linewidth]{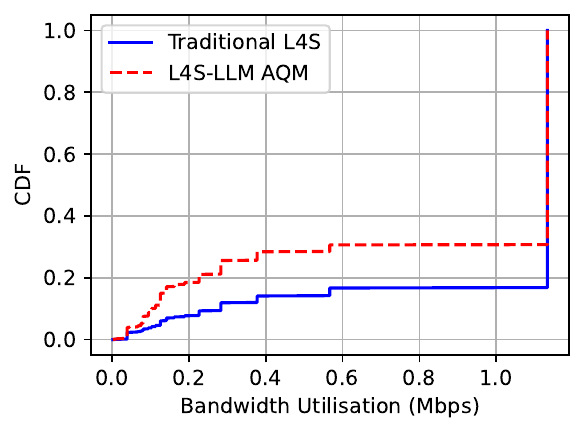}%
        \label{fig:thrptcdf100}
    }
    \caption{CDF of Bandwidth utilization with LLM Enabled at 10 and 100-Packet Intervals.}
    \label{fig:thrptcdfComparison}
\end{figure}

Figure~\ref{fig:thrptboxComparison} and Figure~\ref{fig:bpthrptboxComparison} showcase the comparative analysis of bandwidth utilization of the traditional L4S and L4S-LLM framework. The results show a significant increase in median bandwidth utilization, with significantly low variability, although along with a few large outliers. We found that these outliers mainly appear at the start of the data transfer and are not present after a certain duration. In contrast, the traditional L4S algorithm in Figure~\ref{fig:thrptcdfComparison} shows a low median bandwidth utilization value with high variability. Similarly, L4S-LLM shows its effectiveness in optimizing bandwidth utilization while maintaining low queue delay. 

The CDFs of bandwidth utilization for both traditional L4S and L4S-LLM AQM, as depicted in Figure~\ref{fig:thrptcdfComparison}, offer an alternative visualisation of router performance. These results underscore the LLM's capability to enhance queue management strategies effectively.

\subsection{Computation Overhead(s)}
\begin{figure}[htbp]
\centerline{\includegraphics[width=0.5\linewidth]{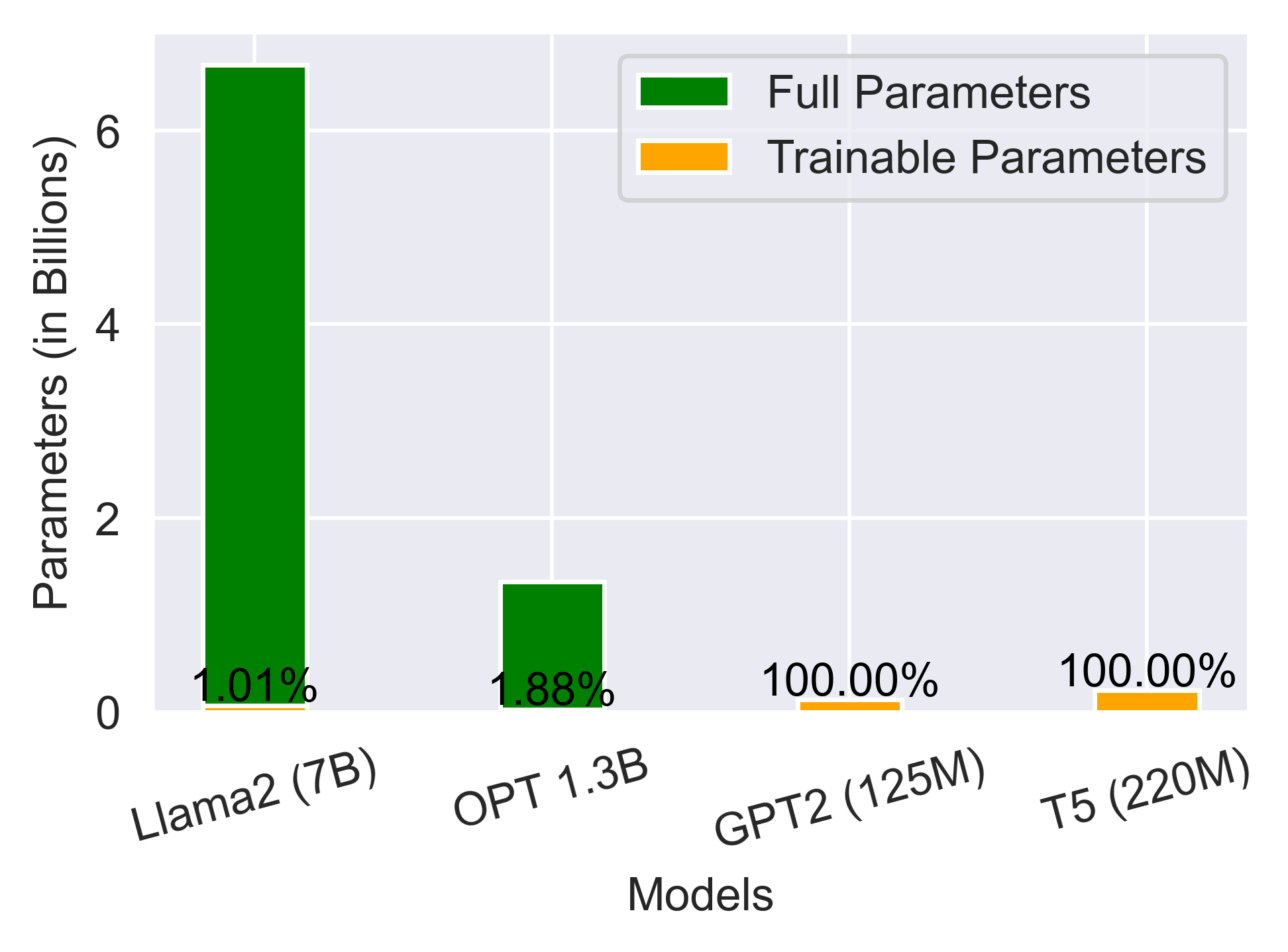}}
 \caption{Distribution of Trainable Parameters Across Four Distinct LLMs}
\label{fig:trainable_parameter_percentage}
\end{figure}

\begin{figure}[htbp]
\centerline{\includegraphics[width=0.5\linewidth]{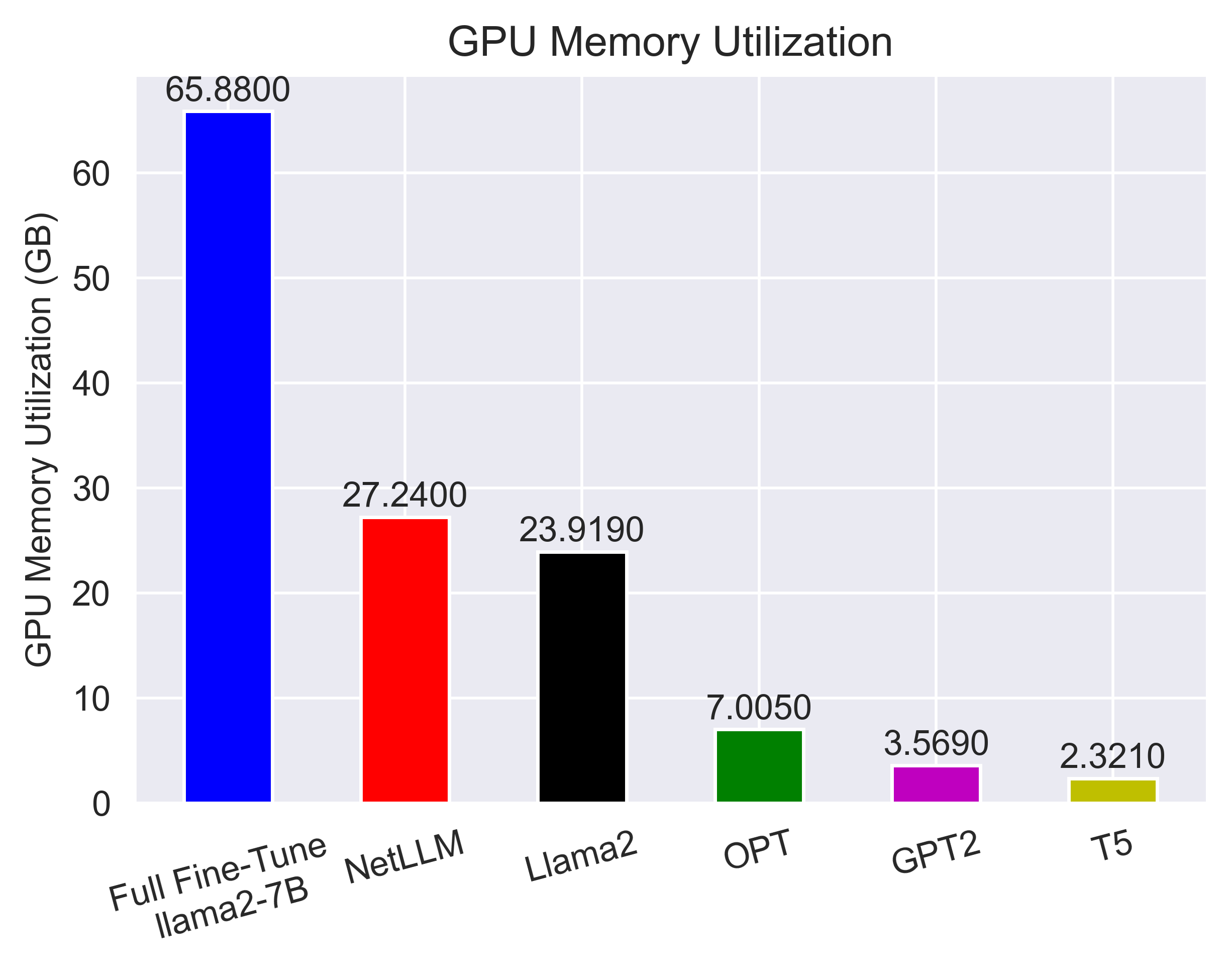}}
 \caption{Comparative Analysis of GPU VRAM Usage Across Different LLMs}
\label{fig:GPU_mem_utilization}
\end{figure}

\color{black}
Figure~\ref{fig:trainable_parameter_percentage} demonstrates LoRA's parameter reduction efficiency: Llama2's 7B parameters reduce to 70M (1\%), while OPT's 1.3B decrease to 25M (1.9\%) using rank-128 LoRA. GPT-2 (125M) and T5-LLM (220M) need no reduction due to their manageable size. This rank-128 configuration optimally balances efficiency with performance, maintaining full fine-tuning accuracy while reducing computational demands. Figure~\ref{fig:GPU_mem_utilization} highlights memory constraints critical for deployment: Llama2 requires 23.91GB VRAM, OPT 7GB, GPT2 3.6GB, and T5-LLM just 2.32GB. These requirements emphasize the need for efficient architectures in resource-constrained networking environments where routers must operate within strict hardware limitations while maintaining real-time performance.

\begin{figure}[htbp]
\centerline{\includegraphics[width=0.5\linewidth]{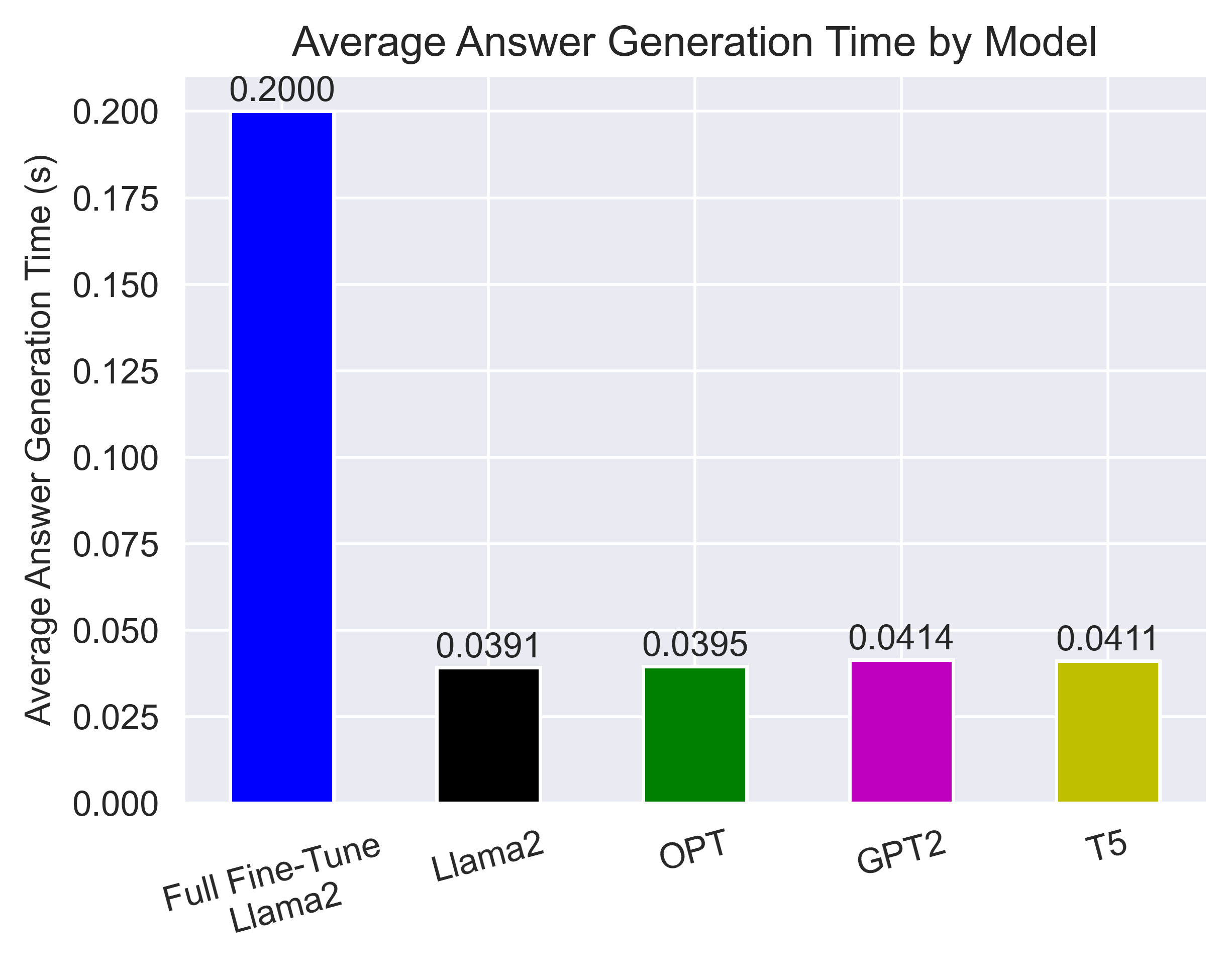}}
 \caption{Comparison of Decision Time Across Different LLMs (s)}
\label{fig:ans_gen_time}
\end{figure}


Figure~\ref{fig:ans_gen_time} demonstrates L4S-LLM's significant impact on inference latency. The unadapted Llama2-7B exhibits prohibitive answer generation time of 0.200 seconds, while LoRA-adapted models achieve substantial reductions: Llama2 at 0.0391s, OPT at 0.0395s, GPT2 at 0.0414s, and T5-LLM at 0.0411s. Despite Llama2-7B and OPT-1.3B containing billions of parameters, LoRA reduces inference time by approximately 80\%, achieving ~0.039s response times crucial for real-time congestion control. This 5x speedup validates L4S-LLM's efficiency for latency-critical network applications. However, further optimization opportunities exist through small language models (SLMs), model pruning, partitioning strategies, and specialized hardware acceleration to achieve sub-millisecond inference suitable for high-frequency packet processing.

\begin{figure}[htbp]
\centerline{\includegraphics[width=0.5\linewidth]{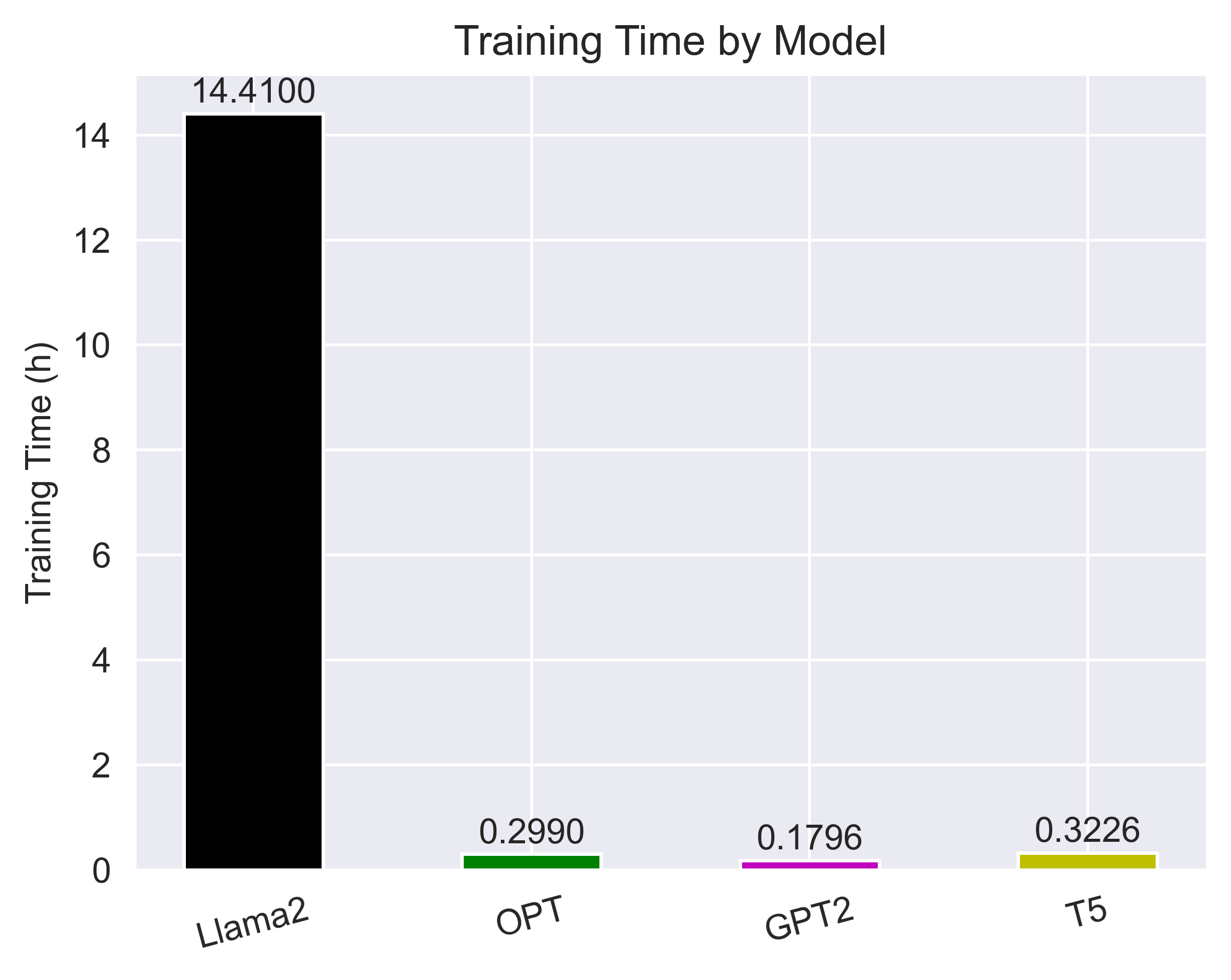}}
 \caption{Comparison of Training Times Across Different LLMs (s)}
\label{fig:train_time}
\end{figure}

In Figure~\ref{fig:train_time}, we observe the total training time required to run 80 epochs for four different LLMs. Even with the application of LoRA, Llama2-7B exhibits the highest training time by a significant margin at approximately 22.3 hours. In contrast, the training time for OPT with LoRA is around 29.9 minutes, while GPT2 requires approximately 17.96 minutes, and T5-LLM requires about 32.26 minutes. With the exception of Llama2, the training times for other LLM models are notably faster. However, during evaluation or actual deployment, the inference time is of utmost importance. Despite Llama2's higher training time, its inference performance is comparable to other models, making it suitable for real-time congestion control applications.

\color{black}
\subsection{Model Architecture and Optimization Considerations}

Our architectural decisions were \textit{strategically formulated} to achieve an optimal balance between performance requirements and deployment feasibility. The L4S-LLM framework implements a robust three-component architecture: (1) a sophisticated state encoder utilizing linear layers for scalar inputs and CNNs for time-series data, (2) a powerful transformer-based reasoning core, and (3) an innovative L4S-LLM head enabling single-inference action generation that significantly reduces decision latency.

Through \textit{extensive network evaluation}, we systematically tested different foundation models (Llama2-7B, OPT-1.3B, GPT2-125M, T5-220M). While Llama2-7B demonstrated superior accuracy (97.56\%), smaller models achieved a substantial 83\% accuracy, providing compelling evidence for viable architectural simplification pathways that maintain acceptable performance levels.

Regarding the critical latency-complexity tradeoff, our measured decision times (39-41ms) currently exceed the ideal 15ms target for high-frequency AQM operations. We effectively mitigated this limitation through periodic execution at 10 and 100-packet intervals (Figures~\ref{fig:qdboxComparison}-\ref{fig:thrptcdfComparison}), conclusively demonstrating that even with reduced execution frequency, L4S-LLM delivers \textit{consistently superior performance} compared to traditional approaches across all key metrics.

Our current implementation focuses on parameter-efficient fine-tuning through LoRA, achieving a \textit{remarkable 99\% reduction} in trainable parameters. We fully recognize the importance of additional optimization techniques for production deployment. Future work will systematically explore structured pruning to eliminate non-essential network components, quantization-aware training for lower precision operation, and hardware-aware optimization specifically targeting router-specific constraints. Based on recent advances in model compression techniques~\cite{guo2025deepseek, 10271124}, these approaches could potentially reduce model size by an additional 75-90\% with minimal performance degradation, making deployment on router-grade hardware an achievable near-term goal.

Our comprehensive open-source FreeBSD-14 implementation establishes a solid foundation for the networking community to collaboratively explore these optimization approaches as we advance toward production-ready systems. Several critical limitations must be addressed before considering the commercial deployment of the proposed L4S-LLM framework in network routers. This section outlines the primary challenges affecting deployability and overall effectiveness:

\begin{itemize}
    \item \textbf{Online Training/Testing:} 
    The framework requires real-time evaluation within the L4S architecture to address deployment challenges. While the current L4S-LLM framework was trained and tested using a data-driven approach with experimental data from L4S DualPI2 AQM and various congestion control algorithms, this process involved offline training and testing. 

    \item \textbf{Time Constraints:} 
    Deployment requires a significant reduction in L4S-LLM's response generation time to ensure effective performance in real-world applications.

    \item \textbf{Hardware:} 
    Approaches for fine-tuning and distilling LLMs rely on high-end hardware, including an RTX 4090 GPU, 128 GB of RAM, and a 13th-generation i9 processor. Commercial network routers typically lack such resources. Therefore, model parameter size must be reduced to address memory constraints, alongside CPU and GPU optimizations. A potential solution involves transitioning to SLMs, potentially by using correlated knowledge distillation (CKD)~\cite{10271124}, to replicate LLM capabilities on resource-limited devices.

    \item \textbf{Energy:} 
    Continuous operation of LLMs requires substantial energy, typically supported by high-performance GPUs and CPUs. The significant energy demands and associated costs pose a challenge to commercial viability. Major technology firms are exploring nuclear-powered energy solutions to meet their LLM needs. 
\end{itemize}

\color{black}

\section{Conclusion}
\color{black}
This research introduces AQM-LLM, a framework that adapts large language models for active queue management within the L4S architecture. Leveraging preemptive ECN signalling and packet dropping, L4S-LLM dynamically controls congestion, optimizes queues, and minimizes latency. Low-rank adaptation cuts memory use by 64\%, while a data-driven reinforcement learning pipeline streamlines fine-tuning. An open-source L4S-LLM platform on FreeBSD-14 supports future research and IETF standardization. While delivering major gains in latency, throughput, and efficiency, further refinement via pruning, quantization, and distillation toward small-language-model-based AQM can enhance scalability and resource efficiency.
\ifCLASSOPTIONcaptionsoff
  \newpage
\fi




\bibliographystyle{IEEEtran}
\bibliography{references.bib}
\color{black}
\section*{Appendix}

\section*{Towards L4S-LLM Stability and Convergence}

We analyze the stability and convergence properties of L4S-LLM by integrating transformer-based language model theory with control-theoretic principles of L4S AQM systems. Considering transformer dynamics and convergence, the self-attention mechanism in L4S-LLM can be approximated as
\begin{equation}
\text{Attention}(Q, K, V) = \text{softmax}\left(\frac{QK^T}{\sqrt{d_k}}\right)V.
\end{equation}
where $Q, K, V \in \mathbb{R}^{n \times d_k}$ represent query, key, and value matrices derived from network state embeddings. For network state sequences $s = (s_1, s_2, \ldots, s_n)$, the attention mechanism creates a weighted representation that captures temporal dependencies across network metrics. The transformer layer mapping $F$ exhibits convergence when its Lipschitz constant satisfies
\begin{equation}
\mathcal{L} = \sup_{h \neq h'} \frac{\|F(h) - F(h')\|}{\|h - h'\|} < 1.
\end{equation}

With LoRA adaptation, the effective weight matrices become $W = W_0 + AB$ where $W_0$ is the pre-trained weight, and $A \in \mathbb{R}^{d \times r}$, $B \in \mathbb{R}^{r \times k}$ are low-rank adaptation matrices. The stability of this adapted system depends on the spectral radius of the Jacobian:

\begin{equation}
\rho\left(\frac{\partial F_{\text{LoRA}}(h)}{\partial h}\right) < 1
\end{equation}
where $F_{\text{LoRA}}$ is the LoRA-adapted transformer mapping. Our empirical results with rank $r=128$ confirm this stability condition is maintained.

Employing L4S control theoretic approach, the L4S-LLM policy $\pi_\theta: \mathcal{S} \rightarrow \mathcal{A}$ replaces the traditional PI controller dynamics
\begin{equation}
p'(t) = p'(t-1) + \alpha(q_{\text{delay}}(t) - q_{\text{target}}) + \beta(q_{\text{delay}}(t) - q_{\text{delay}}(t-1)).\nonumber
\end{equation}
Given Lyapunov function $V(s) = (q_{\text{delay}} - q_{\text{target}})^2$, stability is guaranteed when
\begin{equation}
\mathbb{E}[V(s_{t+1}) | s_t, \pi_\theta(s_t)] - V(s_t) < 0, \quad \forall s_t \neq s^*
\end{equation}
For the RL-optimized policy, convergence follows
\begin{equation}
\|J(\pi_k) - J(\pi^*)\| \leq (1-\gamma\lambda)^k \|J(\pi_0) - J(\pi^*)\|
\end{equation}
where $J(\pi)$ is the expected return, $\pi^*$ is the optimal policy, and $\lambda$ relates to the mixing time of the induced Markov chain. Therefore, L4S-LLM provides probabilistic queue stability
\begin{equation}
\mathbb{P}\left(\sup_{t \geq T_0} |q_{\text{delay}}(t) - q_{\text{target}}| \leq \epsilon\right) \geq 1 - \delta
\end{equation}
for finite $T_0$, stability bound $\epsilon$, and error probability $\delta$. 

Observe that this theoretical analysis is substantiated by the experimental results discussed earlier, for examples, rapid convergence of training loss (Figure~\ref{fig:trainingloss}), high accuracy in action prediction (97.56\% for Llama2), significant improvements in queue delay stability (Figure~\ref{fig:qdboxComparison}), enhanced bandwidth utilization (Figure~\ref{fig:thrptboxComparison}). Our understanding is that the integration of LLM and ha the potential to yield an enhanced AQM system with strong stability and convergence properties, capable of outperforming traditional approaches.

\section*{Model Compressions for Router-Deployable L4S-LLM}
To address L4S-LLM deployment's computational limitations, we propose four optimization strategies to reduce model size while maintaining performance:

\paragraph{Structured Pruning} Using attribution scores and confidence-weighted patterns, approximately 40\% of attention heads could be eliminated with negligible impact, as network traffic patterns exhibit strong temporal locality. Selective pruning of FFNs (65\% of parameters) based on activation patterns during congestion events could reduce dimensions by 50-60\% while maintaining responsiveness. Layer-specific compression rates guided by sensitivity analysis would preserve parameter density in layers with greater impact on queue management.

\paragraph{Quantization-Aware Training} Critical components would maintain FP16 precision, while less sensitive components could operate at INT8/INT4, reducing memory requirements by up to 75\%. Quantization parameters would be calibrated using diverse traffic patterns, emphasizing accuracy during congestion transition states. Adaptive precision would allocate higher bit-width to components processing critical congestion indicators while using lower precision for stable states.

\paragraph{Neural Architecture Search} Evolutionary search algorithms would explore architectural variants optimized for congestion control. Asymmetric encoder designs with wider attention layers but narrower feedforward networks could maintain 95\% performance while reducing parameters by 70\%. Custom components for network telemetry could replace general-purpose transformers, with dedicated convolutional pathways for multi-timescale queue dynamics.

\paragraph{Hardware-Aware Optimization} Critical inference paths would be optimized to reduce memory transfers for efficient execution on processors with limited cache. The model would develop structured sparsity patterns aligned with hardware execution blocks. Router-specific optimizations would include operator reordering, buffer reuse, and execution scheduling tailored to hardware constraints.

\end{document}